\documentclass[twocolumn,secnumarabic,amssymb,nobibnotes,aps,pra,showpacs,superscriptaddress]{revtex4-1}
\usepackage[english]{babel}
\usepackage{graphicx}
\usepackage{amsmath}
\usepackage{amsfonts}
\usepackage[usenames,dvipsnames]{color}
\usepackage{hyperref}
\usepackage{blindtext}
\hypersetup{colorlinks}
\hypersetup{linkcolor=black, citecolor=black, urlcolor=blue}
\usepackage[utf8]{inputenc}
\usepackage{color}
 \usepackage{xspace}
 \newcommand\mycolor{\color{black}\xspace}

\begin{document}
\title{Alternating strings and clusters in suspensions of charged colloids}
\author{J. C. Everts}
\email{j.c.everts@uu.nl}
\address{Institute for Theoretical Physics, Center for Extreme Matter and Emergent Phenomena,  Utrecht University, Princetonplein 5, 3584 CC Utrecht, The Netherlands}
\email{j.c.everts@uu.nl}
\author{M. N. van der Linden}
\thanks{Present Address: Department of Chemistry, Physical and Theoretical Chemistry Laboratory, University of Oxford, South Parks Road, Oxford OX1 3QZ, United Kingdom}
\address{Soft Condensed Matter, Debye Institute for Nanomaterials Science, Princetonplein 5, 3584 CC, Utrecht, The Netherlands}
\author{A. van Blaaderen}
\author{R. van Roij}
\address{Institute for Theoretical Physics, Center for Extreme Matter and Emergent Phenomena,  Utrecht University, Princetonplein 5, 3584 CC Utrecht, The Netherlands}
\date{\today}

\begin{abstract}
We report the formation of alternating strings and clusters in a binary suspension of repulsive charged colloids with double layers larger than the particle size. Within a binary cell model we include many-body and charge-regulation effects under the assumption of a constant surface potential, and consider their repercussions on the two-particle interaction potential. We find that the formation of induced dipoles close to a charge-reversed state may explain the formation of these structures. Finally, we will touch upon the formation of dumbbells and small clusters in a one-component system, where the effective electrostatic interaction is always repulsive.
\end{abstract}

\maketitle

\section{Introduction}
A common theme in soft condensed matter physics and biological physics is the formation of large ordered structures from smaller building blocks \cite{Jones, Alberts} that are sometimes analogs of ordered phases in (hard) condensed matter. These structures can be macroscopic in size, such as bulk (liquid) crystals \cite{Lubensky, Vroege:1992}, plastic crystals \cite{Timmermans:1961, Liu:2014}, quasicrystals \cite{Schechtman:1984, Talapin:2009} and ferrofluids \cite{Genc:2014}, but also mescoscopic structures are found, such as the double-stranded helix in DNA molecules \cite{Watson:1953}, the secondary (and ternary) structures in proteins \cite{Dill:2012}, micelles and membranes \cite{Jones}, and periodic structures in block copolymers \cite{Leibler:1980}.  In this paper, we will show that like-charged colloidal spheres can form also mesoscopic structures, such as alternating strings and clusters, which we will investigate both in experiment and in theory.


In general, charged colloidal particles are interesting due to their highly tunable effective interactions \cite{Russel, Anand:2007}. For example, their charge is often highly adaptable since its microscopic origin is due to the ionic adsorption on or desorption from the colloidal surface, a phenomenon called charge regulation \cite{Ninham:1971}, and there are many possibilities for tuning the dielectric and screening properties of the surrounding medium \cite{Blaaderen:2003}. Not only energetic effects (Coulomb interactions) are thus important, but also the entropic effects of the (adsorbed and bulk) ions, which can be taken into account by integrating them out of the partition sum \cite{Roij:1997, Zoetekouw, Lowen2:1993, Warren:2000, Denton:2005, Zoetekouw:2006}. For low electrostatic potentials, the resulting effective interaction between pairs of colloids is of the Yukawa type, and the strength and range can be tuned by varying temperature, the salt concentration and density of particles. Taken together with the attractive van der Waals force, which can be significantly reduced by matching the dielectric constants of the particles and solvent in the visible range of frequencies \cite{Israelachvili}, the effective pair potential is known as Derjaguin-Landau-Verwey-Overbeek (DLVO) potential \cite{Derjaguin:1948, VerweyOverbeek}. 

Just adding an additional colloidal species can already induce local ordered structures in charged colloidal systems, as was shown in Ref. \cite{Nakamura:2015} for a binary system of positively charged sub-micron polystyrene and silica spheres. These particles could become negatively charged, however, as a function of surfactant concentration due to charge regulation, but the exact concentration for which this occurs is different for the two species. Hence, clusters were observed when only one of the two species became negatively charged, since opposite charges attract. 

If the charge itself is inhomogeneously distributed on the particle surface, it can be shown that the effective interaction can be decomposed in multipoles \cite{Hoffman:2004, Ramirez:2006, Boon:2010, Graaf:2012}, which provide to leading order monopole-monopole (DLVO), monopole-dipole and dipole-dipole interactions. These inhomogeneities can have a large impact on the self-assembly, as we shall see in this paper. The fact that higher order moments of the charge distribution are important, was already hinted towards in Ref. \cite{Smallenburg:2012}. Here a binary mixture of colloidal species that are both positively charged was considered, such that one expects (at least on the mean-field level) that the effective monopole-monopole interaction is repulsive. However, when an external electric field is applied, dipoles are induced on these particles, that can drive the formation of various structures, showing that the dipole-dipole interaction can become more important than the monopole-monopole interaction, see for experimental examples Refs. \cite{Leunissen:2005, Vissers:2011}. Moreover, theoretical work shows that a long-range repulsion, combined with a short-distance attraction, can lead to a regime where microphase separation occurs instead of bulk phase separation \cite{Sear:1999, stradner:2004, Imperio:2004, Archer:2007, Groenewold:2001, Groenewold:2004}, which will be a theme here as well.

An external electric field is not the only way to induce dipole moments in charged colloidal particles. For example, for a binary mixture of oppositely charged particles it is known that dielectric effects can give rise to string formation \cite{Luijten:2014}. In the calculation of Ref. \cite{Luijten:2014}, however, salt was not included, which appears to be a severe approximation in systems where charge regulation plays a significant role. 

Charge-regulating particles usually discharge when the interparticle distance is reduced \cite{Zoetekouw, Vissers2:2011}, which weakens electrostatic repulsion. When the particle surroundings are anisotropic, this will lead to an inhomogeneous charge distribution, and hence to a self-induced dipole moment. The adjective ``self" is used here to make a distinction with dipoles that are, for example, induced by an external electric field. Self-induced dipoles will turn out to be important for the formation of alternating strings and clusters in a binary system of colloidal particles, which we report here. Our particles are positively charged for one-component suspensions in the low-polar solvent cyclohexylbromide (CHB) that we use, as was shown earlier by electrophoresis measurements  \cite{Linden:2015}. In low-polar solvents the electrostatic interactions can play an even more important role than in water, since the Coulomb interactions are stronger (dielectric constant is lower) and the screening lengths are longer \cite{Morrison:1993, Blaaderen:2003}.  We will show that a combination of charge regulation and an asymmetry in the charge distribution of both species can lead to a short-distance attraction on top of the long-range repulsion. This rationalizes the formation of the observed strings, while more compact clusters are formed when one of the two species becomes (almost) negatively charged in bulk. We close this paper with the observation of dumbbells in one-component systems, that were also earlier observed \cite{vdLinden:2013} and for which self-induced dipoles do not seem to give an explanation.
\section{Experimental method}
\label{sec:experimental}
We prepared dispersions of poly(methyl methacrylate) (PMMA) colloidal spheres in cyclohexyl bromide (CHB), sterically stabilized by a comb-graft copolymer of poly(12-hydroxystearic acid) (PHSA) grafted onto a PMMA backbone. The colloids (red $r$ and green $g$ in this work) had radii $a_r=0.79\ \mu\mathrm{m}$ and $a_g=0.99 \ \mu\mathrm{m}$, and were labeled by the fluorescent dyes rhodamine isothiocyanate and 7-nitrobenzo-2-oxa-1,3-diazol, respectively. This system is the same as sample 6 in Ref. \cite{Linden:2015}, meaning that the green particles underwent a locking procedure, and hence have a higher charge than the red particles \cite{Linden:2015}. We performed electrophoresis measurements on this binary mixture one day after mixing. We found that the measured surface potentials at volume fractions $\eta_r=\eta_g=0.028$ are $\phi_g=6.5\pm0.1$ and $\phi_r=3.29\pm0.09$ (in units of $k_B T/e$, with $k_BT$ the thermal energy and $e$ the proton charge), with corresponding charges $Z_g=1015\pm 40$ and $Z_r=153\pm9$ in units of $e$. Although for the conversion from mobility profiles to these quantities a one-component theory is used, we do believe that this procedure gives rather realistic results. Further experimental details can be found in Ref.\ \cite{Linden:2015}.
\section{Formation of alternating strings and clusters}
\label{sec:form}
\begin{figure*}[ht]
\includegraphics[width=0.75\textwidth]{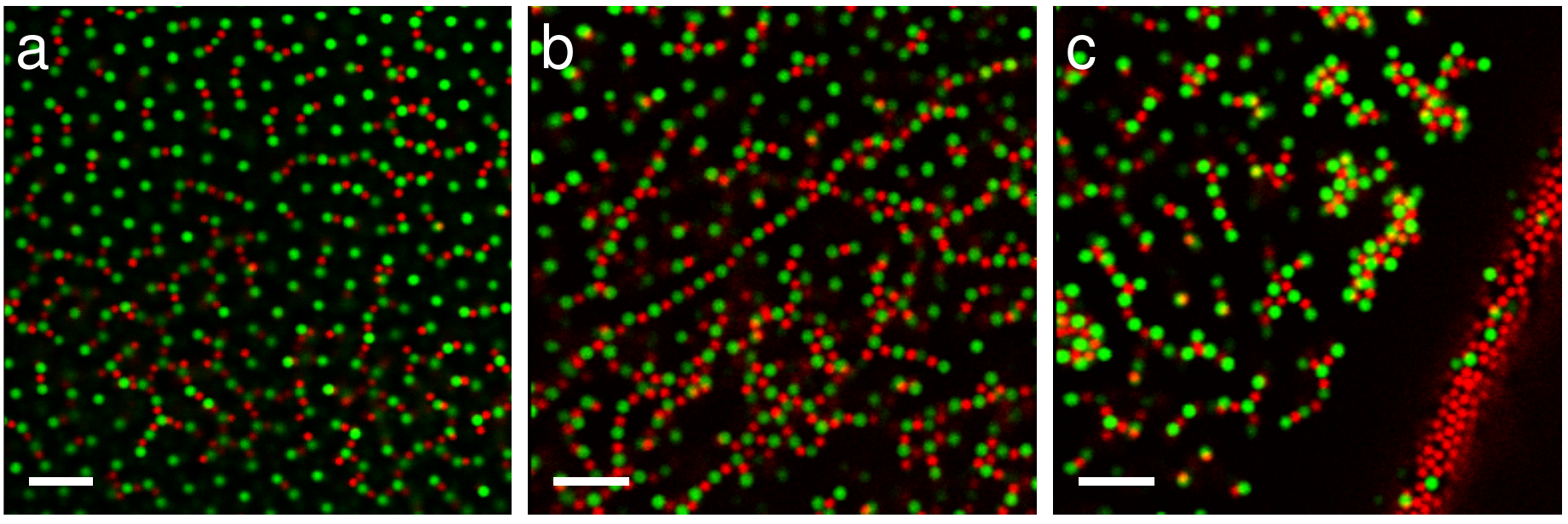}
\caption{Alternating strings formed in binary suspensions of large (green, with radius $a_g=0.99\ \mu\mathrm{m}$) and small (red, with radius $a_r=0.79\ \mu\mathrm{m}$) particles in CHB (a) without deionised water; two days after mixing, (b) in contact with deionised water; 0.2 mm from CHB-water interface, $\sim 6.5$ hours after mixing, and (c) where extended clusters are found near the CHB-water interface; $\sim 4.5$ hours after mixing. Particles are adsorbed to the CHB-water interface, making the interface clearly visible; the water phase is on the right, the CHB phase, containing the clusters, is on the left. Scale bars indicate 10 $\mu$m. Movies are available for (a)-(c) as movie 1-3, respectively, in the supplementary information.}
\label{CHfig:str}
\end{figure*}

\begin{figure*}[ht]
\includegraphics[width=0.8\textwidth]{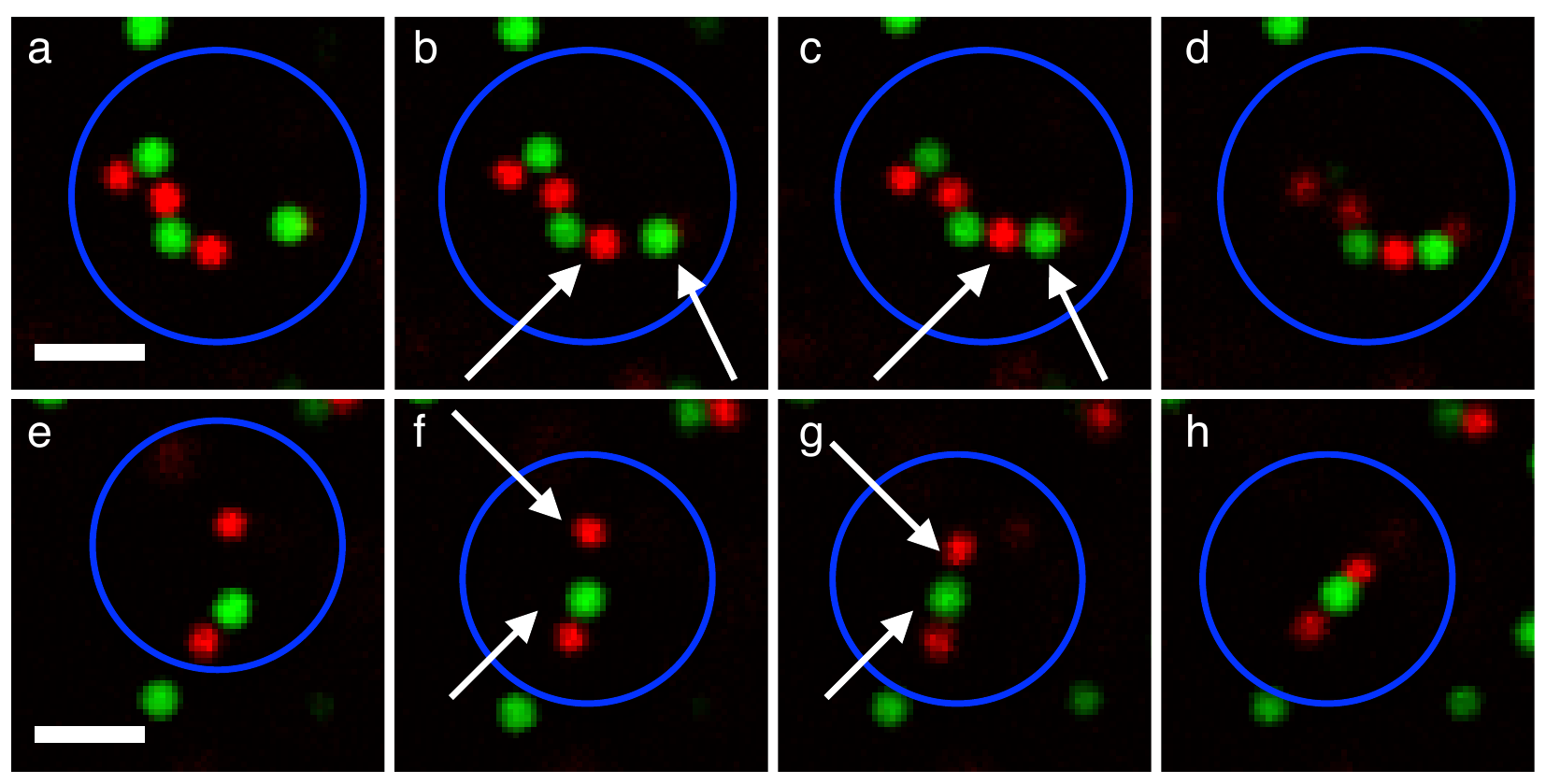}
\caption{Two sequences of four confocal images (a)--(d) and (e)--(h) showing the attachment of a large (green) and small (red) particle. The images were taken 1.3 mm and 1.1 mm from the CHB-water interface, $\sim 1$ hours and $\sim 4$ hours after mixing, respectively. In each case we show, from left to right, two images that were taken before attachment, followed by two images taken after attachment. The time step between two consecutive images was 1.04 s and the scale bars indicate 5 $\mu$m. An uncropped version of this Figure is available in the Supplementary information \cite{supp}. The corresponding movies (movie 4 for (a)-(d) and movie 5 for (e)-(h), uncropped) are also available in the supplementary information \cite{supp}. In these movies the attachments shown in this figure occur after 5 seconds.}
\label{CHfig:att}
\end{figure*}

When checking a sample that had been prepared two days earlier, we observed strings that consist of alternating green and red particles (see Fig.\ \ref{CHfig:str}(a)). The strings consisted almost exclusively of this alternating sequence: red particles had green neighbours and green particles had red neighbours. Sometimes we encountered a red particle flanked by a green and a red neighbour. We ascribe this to the incorporation of a pre-existing red-red dumbbell into the string, as we occasionally also observed free red-red dumbbells in the binary suspensions. 
The strings were not rigid: the bond angles in a string were still fluctuating{\mycolor (see movies in} the supplementary material \cite{supp}). From the image in Fig.\ \ref{CHfig:str}(a) it is clear that the interactions between particles in a string and free particles are still long-range repulsive: the green particles stay far apart.

One explanation for the occurrence of these alternating strings could be that the two species became oppositely charged when mixed together. To check this, we performed electrophoresis measurements on a binary mixture in which a few green-red clusters had formed (one day after mixing), however, we saw no indication that the particles had acquired net charges of opposite sign. We take the presence of a few green-red clusters as an indication that the interactions in this measured sample were close to those in the sample where extensive string formation took place (two days after mixing). As there were far more single particles than clusters, the clusters presumably had a negligible effect on the measured mobility profiles.
Thus, the system contained two species of like-charged (positively charged) particles, which spontaneously formed alternating strings (alternating meaning here that green particles had red neighbours and vice versa). From the electrophoresis measurements on the one-component suspensions  and the mixture \cite{Linden:2015}, we know that the green particles had a higher charge than the red particles. We hypothesize that a higher-charged green particle, upon approaching a lower-charged red particle, could induce a patch of opposite charge on the red particle, inducing a dipole moment on the particle, which then caused the green and red particle to attach. This hypothesis will be investigated quantitatively below.

We also prepared a sample containing the binary suspension of red and green particles adjacent to a deionised water phase, which is expected to act as a sink for the ions. Surprisingly, the presence of the water phase strongly promoted the formation of extended alternating strings and clusters (see Figs.\ \ref{CHfig:str}(b) and (c)). Close to the water phase (Fig.\ \ref{CHfig:str}(c)) we observed ``alternating'' clusters that were more compact than strings, yet still quite extended. Further from the water phase we observed a network of alternating strings (Fig.\ \ref{CHfig:str}(b)). These strings appeared less flexible than in the sample without water (Fig.\ \ref{CHfig:str}(a)), see movies 1 and 2 in the supplementary information \cite{supp}. At other places in the capillary, below and above the network of strings, we encountered shorter strings, small clusters and single green and single red particles (see Fig. S1 for typical images).
At this point, we do not know the mechanism behind the enhanced formation of clusters in the presence of water. Previous work showed that the water phase may take up ions from the CHB phase \cite{Leunissen:2007}, reducing the ionic strength in the suspension and thus changing the particle interactions. A lower ionic strength corresponds to a larger Debye length $\kappa^{-1}$ and may lead to a lower charge on the particles through particle discharging \cite{Smallenburg:2011}. It seems unlikely that water itself plays a role in string formation, as we also observed clusters and strings in suspensions made of purified CHB {\mycolor(from which water, ions and other (polar) chemicals were removed following the procedure of Ref. \cite{Linden:2015})} without any contact with a water phase (Fig.\ \ref{CHfig:str}(a)). 
\begin{figure*}[ht]
\includegraphics[width=0.9\textwidth]{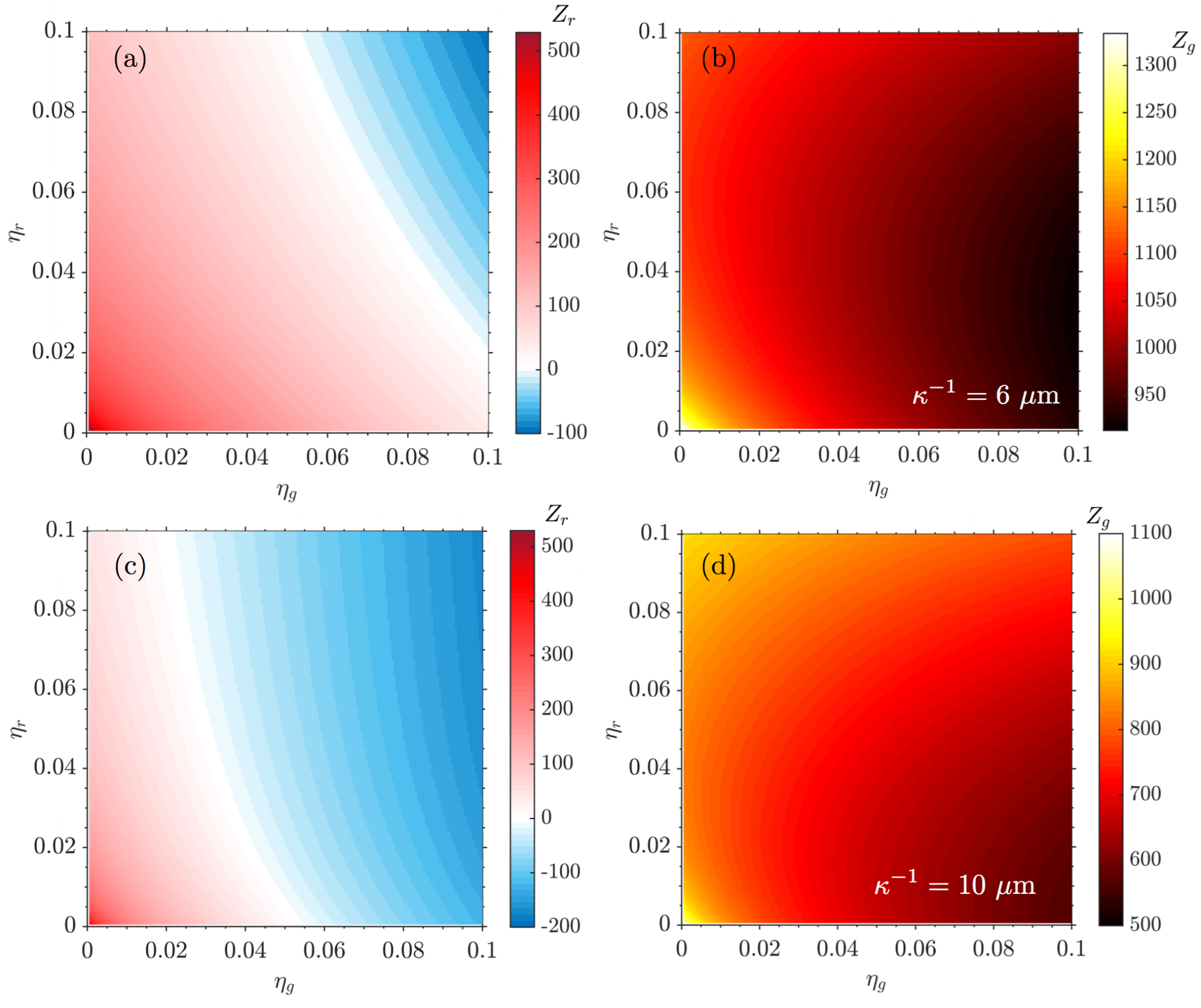}
\caption{Colloidal charges for the (a) red particles $Z_r$ using a constant surface potential of $\phi_r=4.4$ and (b) green particles $Z_g$ using a constant surface potential of $\phi_g=6.7$ at $\kappa^{-1}=6\ \mu\mathrm{m}$ and at a lower ionic strength of $\kappa^{-1}=10\ \mu\mathrm{m}$ in (c) and (d), respectively. We see that the lower charged particle can become negatively charged at sufficiently high packing fractions of both red ($\eta_r$) and green ($\eta_g$) particles, as indicated in blue in (a) and (c). The green particle is always positively charged as can be seen in (b) and (d).}
\label{fig:charges}
\end{figure*}

Further information on the mechanism of cluster formation could come from analysis of the attachment processes. We succeeded in capturing a few of the rare attachment events with confocal microscopy. We prepared a sample containing the binary suspension and deionised water and put this vertically, with the water phase on top and gravity pointing along the length of the capillary. The particles (having a lower density than CHB) sedimented towards the CHB-water interface. We recorded movies of $\sim$ 1000 frames at a rate of $\sim$ 1 frame per second. By careful inspection of the movies we were able to find a few attachment events where a green particle and a red particle approached each other and attached. 
Fig.\ \ref{CHfig:att} shows two series of images depicting such attachment events. In the first series \mbox{(Figs.\ \ref{CHfig:att}(a)--(d))} a green member of a green-red dumbbell attached to a red particle, which is part of a alternating string. Note that the red member of the green-red dumbbell is not visible in frames (a) and (b) and only barely visible in frames (c) and (d). 
In the second series (Figs.\ \ref{CHfig:att}(e)--(h)) a single red particle attaches to a green particle, which is part of a green-red dumbbell. In both cases the white arrows indicate the two attaching particles in the frame right before and right after attachment. Dynamically, the attachment was simply visible as a transition from two particles moving rather independently of each other to two particles moving coherently and staying at short distance from each other (see movies 4 and 5 in the supplementary information \cite{supp}). A more detailed investigation of these events, and more statistics, would be needed to better characterize the cluster and string formation experimentally, but this was beyond the experimental possibilities.

\section{Spherical cell approximation}
\label{sec:bincell}

Before we rationalize the formation of the alternating strings theoretically, we will first investigate the charge regulation properties of the binary suspension discussed in Section \ref{sec:form}. For this we use the spherical-cell approximation \cite{Alexander:1984, Tamashiro:1998, Trizac:2002, Eggen:2009, Denton:2010}: every colloid of type $i=r,g$ is situated in a spherical cell of type $i$ with radius $R$, and the cells fill the whole system volume, such that $4\pi R^3/3=V/(N_r+N_g)$, with $N_i$ the particle number of the respective species and $V$ the volume. Consequently, we may relate the volume fraction $\eta_i=N_i(4\pi/3)a_i^3/V$ to the number fraction $x_i=N_i/(N_r+N_g)$ via $\eta_i=x_i(a_i/R)^3$. The ionic density profiles around a particle of species $i$ are described by the Boltzmann distributions $\rho_\pm^i(r)=\rho_s\exp[\mp\phi_i(r)]$, with $2\rho_s$ the ion concentration in a (hypothetical) ion reservoir in osmotic contact with the suspension. Together with the Poisson equation for $\phi_i(r)$, it gives the spherically symmetric Poisson-Boltzmann equations 
\begin{equation}
\phi_i''(r)+\frac{2}{r}\phi_i'(r)=\kappa^2\sinh[\phi_i(r)], \quad r\in[a_i,R],
\end{equation}
with $\kappa^{-1}=(8\pi\lambda_B\rho_s)^{-1/2}$ the Debye length and $\lambda_B=e^2/4\pi{\mycolor{\epsilon_0}}{\mycolor{\epsilon_s}} k_BT$ the Bjerrum length, where ${\mycolor{\epsilon_0}}$ is the permittivity in vacuum, and ${\mycolor{\epsilon_s}}$ the relative dielectric constant of the oily medium. {\mycolor Throughout this paper, we will use the value for CHB, $\epsilon_s=7.92$. Naively, one would argue based on this number that ion-ion correlations are important in this system. However, because the ion density is low ($\sim 10^{-10}$M), we are still in the ideal-gas regime, and therefore Poisson-Boltzmann can be applied without even the need of including Bjerrum pair formation (see, for example Fig. 4 in Ref. \cite{Valeriani:2010}).}

For the boundary conditions, we denote $x=x_g$, and express global charge neutrality by $x\phi_g'(R)+(1-x)\phi_r'(R)=0$. Note that in our formulation a single cell is not necessarily charge neutral, however, the weighted average of two types of cells is neutral, in contrast to the approach in Ref. \cite{Torres:2008}. This has the added advantage that no minimization to the cell radii is needed and that numerical problems are circumvented when the colloids tend to be oppositely charged. Moreover, we impose the continuity condition $\phi_r(R)=\phi_g(R)\equiv\phi_D$, with $\phi_D/\beta e$ the Donnan potential that is found self-consistently. Finally, we also impose constant-potential boundary conditions: $\phi_i(a_i)=\phi_i$, and by applying Gauss' law, we can evaluate the charge $Z_i=-a_i^2\phi_i'(a_i)/\lambda_B$. {\mycolor Note that in the constant-potential approximation the colloidal particles resemble active capacitors because they adjust their charge to ensure that their surface potential is constant. Furthermore, assuming a constant potential is a very good approximation for a charge-regulation model for which cations and anions can simultaneously adsorb on the colloidal surface \cite{Everts:2016}, which is thought to be relevant for the types of particles that we consider, see Ref. \cite{Linden:2015}.}

First, we tried to use the measured values of the surface potentials  $\phi_g=6.5\pm0.1$ and $\phi_r=3.29\pm0.09$ at $\eta_r=\eta_g=0.028$ to determine the dependence of $Z_r$ and $Z_g$ on overall packing fraction and composition. For this condition, we found that $Z_r<0$, while it was established that all species in the suspension are positively charged at this state point. For this reason, we instead use the experimentally measured charges $Z_r=153$ and $Z_g=1015$ at $\eta_r=\eta_g=0.028$ as known quantities  and determine the surface potentials that gave the best correspondence, yielding $\phi_g=6.7$ and $\phi_r=4.4$, which is in reasonable agreement with the experimentally determined zeta potentials.

Using these values of $\phi_r$ and $\phi_g$ for all state points we determine $Z_r$ and $Z_g$ as a function of $\eta_r$ and $\eta_g$ from the binary cell model. The resulting $Z_r$ and $Z_g$ are presented in Fig. \ref{fig:charges}, in (a) and (b) for $\kappa^{-1}=6 \ \mu\mathrm{m}$, and  in (c) and (d) for $\kappa^{-1}=10 \ \mu\mathrm{m}$. We see that the green particle has always a relatively high positive charge, while the red particle can discharge appreciably and even become negative if $\eta_r$ and/or $\eta_g$ are tuned to higher values. Moreover, when the ionic strength is reduced, there is a larger region for which $Z_r<0$. 

\begin{figure}[t]
\includegraphics[width=0.5\textwidth]{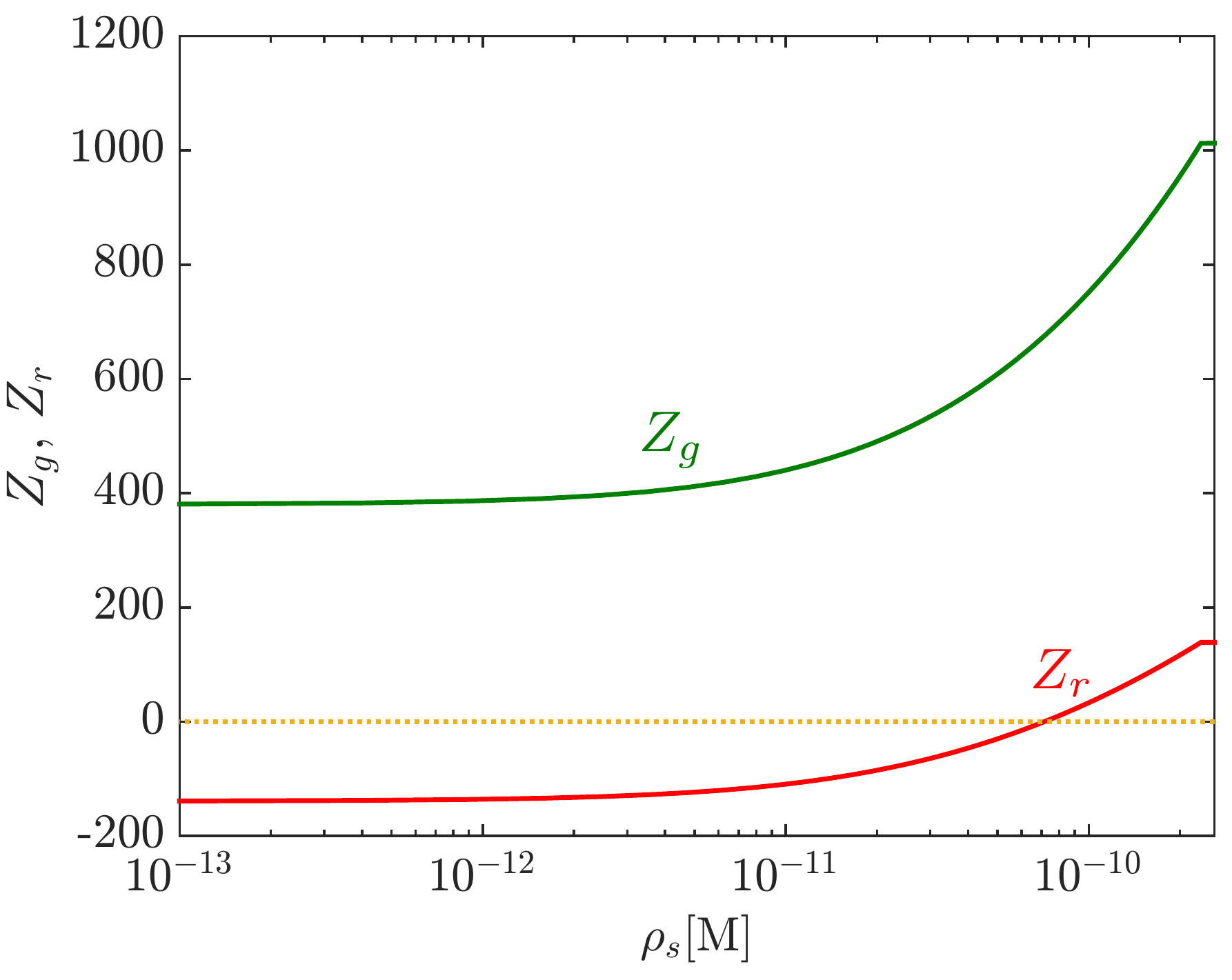}
\caption{Colloidal charges for green particles $Z_g$ (radius $a_g=0.99 \ \mu\mathrm{m}$) and red particles $Z_r$ (radius $a_g=0.79 \ \mu\mathrm{m}$) as a function of salt concentration $\rho_s$ at packing fractions $\eta_r=\eta_g=0.028$ and constant surface potentials $\phi_g=6.7$ (green particles) and $\phi_r=4.4$ (red particles). The dotted line indicates the line $Z=0$. Highest plotted value of $\rho_s$ corresponds with $\kappa^{-1}=6\ \mu\text{m}$.}
\label{fig:chargefunctionofrhos}
\end{figure}

For the experimentally reported packing fractions $\eta_r=\eta_g=0.028$, we show in Fig. \ref{fig:chargefunctionofrhos} how the charges $Z_r$ and $Z_g$ vary with ionic strength. We see that charge inversion occurs at $\rho_s\approx 7\cdot10^{-11}\ \mathrm{M}$, which is about three times lower than the initial salt concentration $\rho_s\approx 2\cdot 10^{-10}\ \mathrm{M}$. This inversion can also be achieved at a higher $\rho_s$ for larger $\eta_r$ and $\eta_g$, as can be deduced from Fig. \ref{fig:charges}. 

\section{Two-body approximation}
\label{sec:twobody}
\begin{figure*}[ht]
\includegraphics[width=\textwidth]{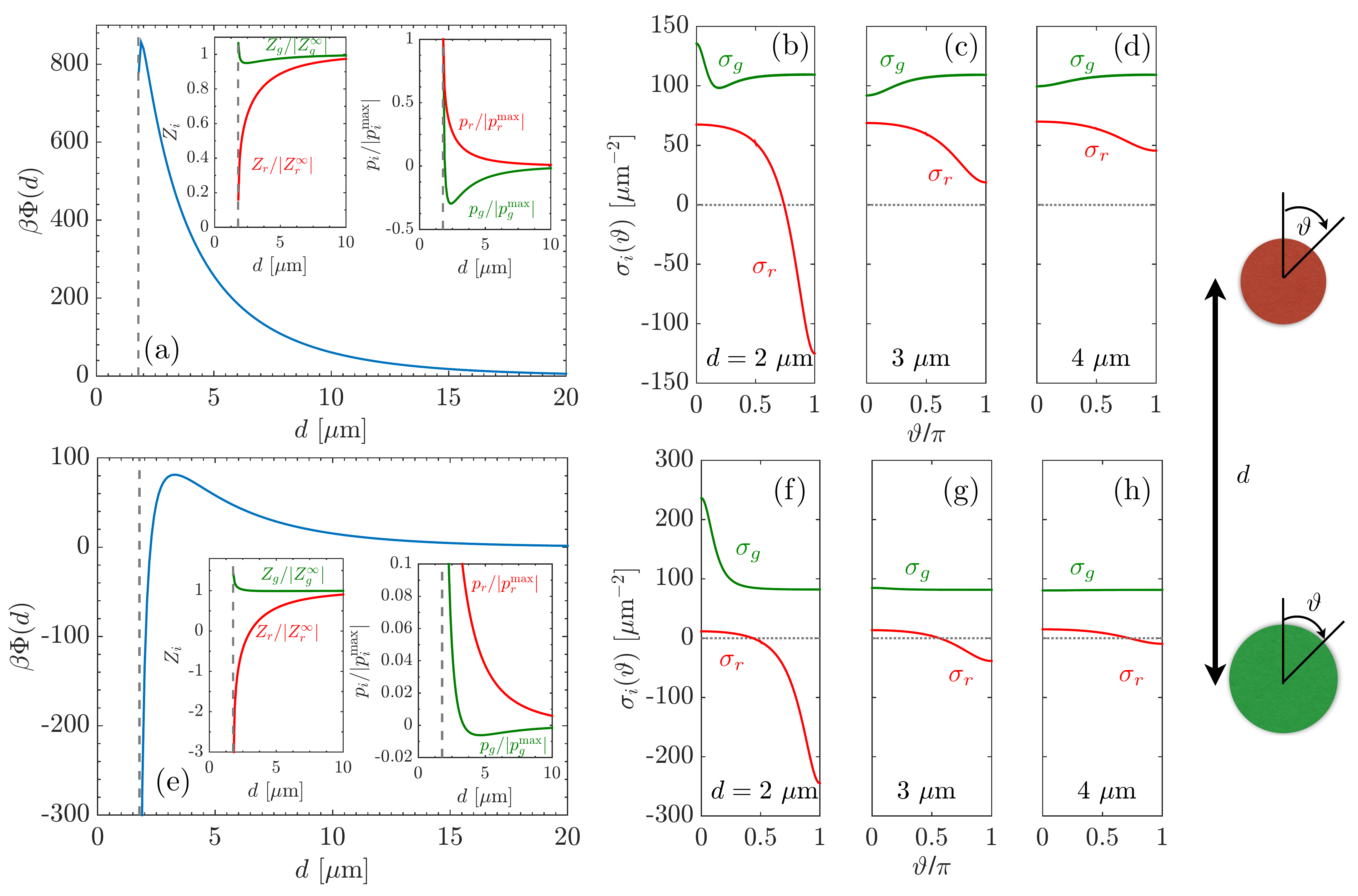}
\caption{Interaction potentials between a red and green particle of radius $a_r=0.79\ \mu\mathrm{m}$ and $a_g=0.99\ \mu\mathrm{m}$, respectively, at Debye lengths $\kappa^{-1}=6 \ \mu\mathrm{m}$ as a function of their center-to-center distance $d$ for (a) surface potentials $\phi_g=6.7$ and $\phi_r=4.4$ and (e) $\phi_g=5.5$ and $\phi_r=1.2$. In the left insets we plot the total charge $Z_i$ ($i=r,g$) normalized by the absolute value of the charge $Z_i^\infty$ at $d\rightarrow\infty$. In (a) we have $Z_g^\infty=1356$ and $Z_r^\infty=582$ , while in (e) we have $Z_g^\infty=1005$ and $Z_r^\infty=152$ for the full lines. In the right inset we plot the dipole moments $p_i$ normalized by the absolute value of the maximum dipole moment $p_i^\text{max}$. The corresponding charge distributions are plotted in (b)-(d) for $\phi_g=6.7$ and $\phi_r=4.4$ and in (f)-(h) for $\phi_g=5.5$ and $\phi_r=1.2$. The dashed vertical grey line in (a) and (e) indicate the distance below which hard-sphere repulsion sets in, while the horizontal dashed grey line in (b)-(d) and (f)-(h) indicates $\sigma_i(\vartheta)=0$.} 
\label{fig:twobody}
\end{figure*}

The spherical symmetry in the cell-model calculations of the previous section cannot directly explain the formation of the alternating strings, since this requires directionality that breaks the spherical symmetry. The simplest extension is the two-body problem, which was investigated earlier within linear Poisson-Boltzmann \cite{Carnie:1994} and later within non-linear Poisson-Boltzmann theory \cite{Stankovich:1996}. Therefore, we now consider two colloids with radii $a_i$, surface potentials $\phi_i$ and centre-of-mass coordinates ${\bf R}_i$ ($i=r,g$) that are a distance $d=|{\bf R}_r-{\bf R}_g|$ apart, dispersed in a solvent with monovalent positive and negative ions described by the density profiles $\rho_\pm({\bf r})$. The solvent is again assumed to be in osmotic contact with a reservoir characterized by a total ion concentration $2\rho_s$, with Bjerrum length $\lambda_B$ and Debye screening length $\kappa^{-1}$.  When we denote the region outside the two colloids by $\mathcal{R}$, the system is described within mean-field theory by the grand potential density functional
\begin{align}
\beta\Omega&[\rho_\pm;d]=\sum_{\alpha=\pm}\int_\mathcal{R}d^3{\bf r} \ \rho_\alpha({\bf r})\left[\ln\left(\frac{\rho_\alpha({\bf r})}{\rho_s}\right)-1\right] \nonumber \\
&+\frac{1}{2}\int_\mathcal{R}d^3{\bf r}\ Q({\bf r})\phi({\bf r})-\sum_{i=1}^2\phi_i\int_{\Gamma_i} d^2{\bf r}\ \sigma_i({\bf r}),
\label{eq:dft}
\end{align}
with the total charge density (in units of e) $Q({\bf r})=\rho_+({\bf r})-\rho_-({\bf r})+\sum_{i\in\{r,g\}}\sigma_i({\bf r})\delta(|{\bf r}-{\bf R}_i|-a_i)$ and $e\sigma_i({\bf r})$ the colloidal surface charge density of sphere $i$, and $\phi({\bf r})$ the dimensionless electrostatic potential. Note that we performed a Legendre transformation to an ensemble where the surface potentials $\phi_i$ are fixed, as we will consider constant-potential boundary conditions below. The Euler-Lagrange equations $\delta\Omega/\delta\rho_\pm({\bf r})=0$ yield the equilibrium density profiles $\rho_\pm({\bf r})=\rho_s\exp[\mp\phi({\bf r})]$ for ${\bf r}\in\mathcal{R}$. Together with the Poisson equation for the electrostatic potential, this results in the non-linear Poisson-Boltzmann equation
\begin{equation}
\nabla^2\phi({\bf r})=\kappa^2\sinh[\phi({\bf r})], \quad {\bf r}\in\mathcal{R}, \label{eq:PB}
\end{equation}
together with the constant-potential boundary conditions supplied for the colloidal surfaces $\Gamma_i$ for $i=r,g$,
\begin{equation}
\phi({\bf r})=\phi_i, \quad {\bf r}\in\Gamma_i \label{eq:CP}.
\end{equation}
Far from the colloids we assume that the electric field vanishes, and inside the colloid the Laplace equation $\nabla^2\phi({\bf r})=0$ is to be satisfied. From the solution of the closed set of Eq. \eqref{eq:PB} and \eqref{eq:CP}, it is then straightforward to obtain the charge densities $e\sigma_i({\bf r})$ by evaluating
\begin{equation}
{\bf n}\cdot[\epsilon_c\nabla\phi|_\text{in}-{\mycolor{\epsilon_s}}\nabla\phi|_\text{out}]/{\mycolor{\epsilon_s}}=4\pi\lambda_B\sigma_i({\bf r}), \quad {\bf r}\in\Gamma_i, \label{eq:evalcharge}
\end{equation}
with $\bf n$ the outward-pointing unit surface normal, and $\epsilon_c=2.6$ the dielectric constant of the colloid (PMMA). {\mycolor Note that for the constant-potential boundary condition that we use, the first term in the square brackets in Eq. \eqref{eq:evalcharge} vanishes.}
The effective interaction Hamiltonian of the colloidal pair can then be found by
\begin{equation}
H(d)=\varphi_\text{HS}(d)+\min_{\rho_\pm}\Omega[\rho_\pm;d],
\end{equation}
where the first term is the bare non-electrostatic colloid-colloid potential, for which we take a hard-sphere potential with contact distance $a_r+a_g$, while the last term is the Legendre-transformed equilibrium ionic grand potential. This leads to
\begin{align}
\beta &H(d)=\beta\varphi_\text{HS}(d)-\frac{1}{2}\sum_{i=1}^2\phi_i\int_{\Gamma_i} d^2{\bf r}\ \sigma_i({\bf r}) \nonumber \\
&+\rho_s\int_\mathcal{R} d^3{\bf r}\Big\{\phi({\bf r})\sinh\phi({\bf r})-2[\cosh{\phi(\bf r)}-1]\Big\}.
\end{align}
Finally, we calculate the total charge $Z_i$ and the dipole moment ${\bf p}_i$ of a colloidal particle with respect to ${\bf R}_i$, defined by
\begin{align}
Z_i=&\int_{\Gamma_i}d^2{\bf r} \ \sigma_i({\bf r}), \nonumber \\
{\bf p}_i=&\int_{\Gamma_i}d^2{\bf r} \ ({\bf r}-{\bf R}_i)\sigma_i({\bf r}),\quad i=1,2.
\end{align}
Note that the definition of the dipole moment requires a specific point of reference that we set to ${\bf R}_i$, because ${\bf p}_i$ is only independent of the reference point for charge neutral particles.

In Fig. \ref{fig:twobody} we plot (a) the interaction potential $\Phi(d)=H(d)-H(\infty)$ for the surface potentials $\phi_g=6.7$ and $\phi_r=4.4$ that were determined from the cell calculations, and we also show the charge $Z_i$ (left inset in (a)), the dipole moment ${\bf p}_i=p_i\hat{\bf z}$ (right inset (a)), where $\hat{\bf z}$ is the unit vector pointing in the same direction as ${\bf R}_r-{\bf R}_g$, and the charge distributions $e\sigma_i(\vartheta)$ for various separations $d$ ((b)-(d)). As can be seen in Fig.  \ref{fig:twobody}(a), $\Phi(d)$ is repulsive for $d\gtrsim 2 \ \mu\mathrm{m}$ at these parameter values, which is reflected also in the equal sign of the charges $Z_r$ and $Z_g$ for both red and green species  for all $d$ and the anti-aligned induced dipole moments (although the red particle does acquire a small negative charge density at its southpole for small enough $d$, which can be seen in Fig. \ref{fig:twobody}(b)). Close to the point where both particles touch, $d\lesssim 2 \ \mu\mathrm{m}$, we see an attraction that is accompanied by the alignment of the induced dipoles, while the particles remain positively charged. 


\begin{figure}[t]
\includegraphics[width=0.5\textwidth]{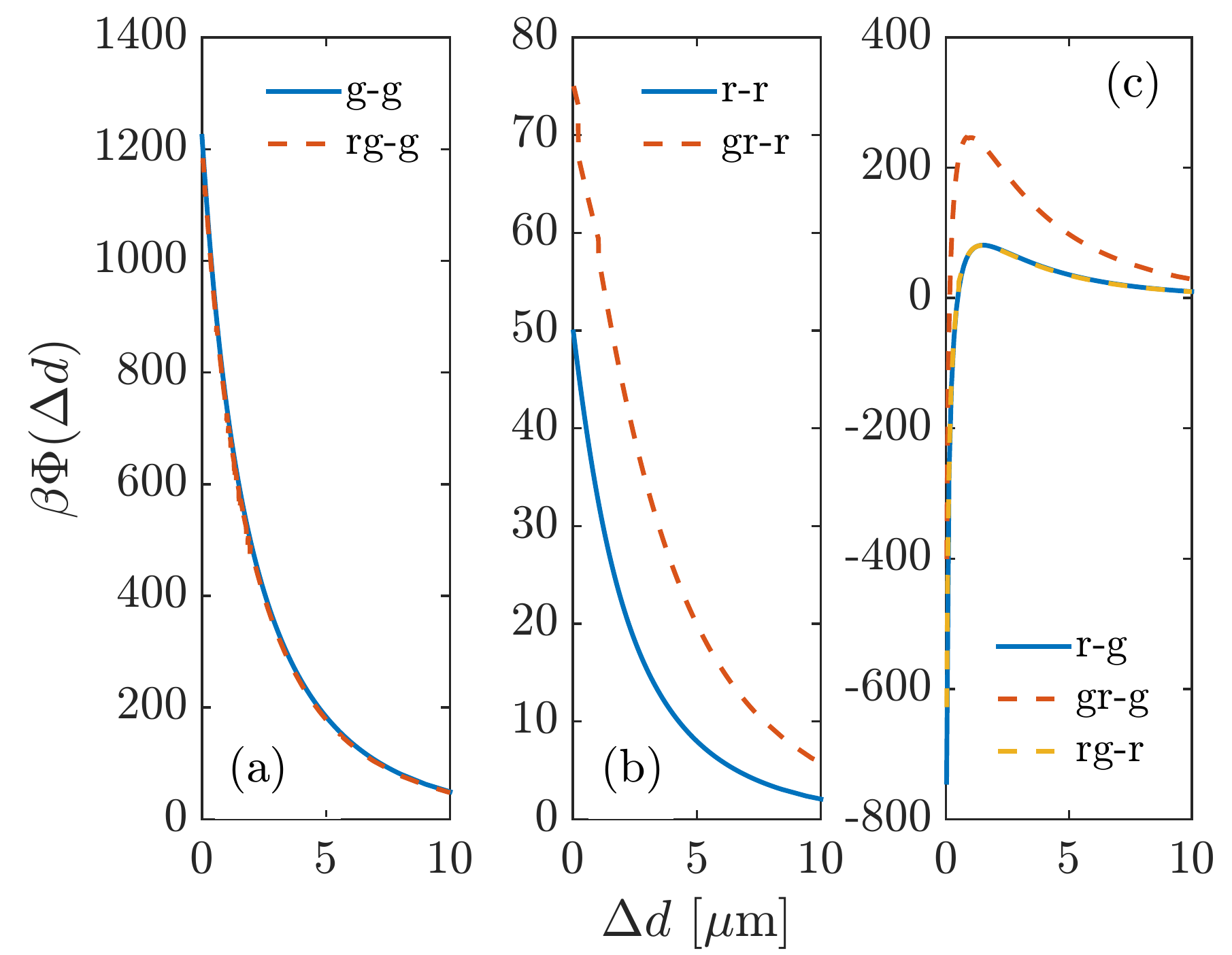}
\caption{Effective interaction potentials $\Phi$ as a function of surface-to-surface distance $\Delta d$ for various configurations of red particles, with surface potential $\phi_r=1.2$ and radius $a_r=0.79 \ \mu\mathrm{m}$ and green particles with surface potential $\phi_g=5.5$ and radius $a_g=0.99\ \mu\mathrm{m}$ at Debye length $\kappa^{-1}=6\ \mu\mathrm{m}$. In (a) we show as the dashed orange line the interaction between a red-green dumbbell and a green particle approaching the green part of the dumbbell. We abbreviate this by rg-g. We compare this effective interaction with the green-green (g-g) repulsion as the full line. Similarly, we show in (b) the r-r and gr-r repulsion and in (c) the effective interactions r-g, gr-g and rg-r that feature short-range attraction and long-range repulsion.}
\label{fig:dumbbelint}
\end{figure}

However, this calculation does not include many-body effects. This can be seen from the observation that for $d$ large the particles have charges $Z_r^\infty=582$ and $Z_g^\infty=1356$ which are larger than the charges determined from the cell model at the experimentally measured packing fractions, and these charges are even much higher when $d$ is on the order of the inter-particle distance $d=3-5 \ \mu\mathrm{m}$. To include the fact that particles discharge when the density is increased, we set the charges of the pair at infinite distance to coincide with the charges determined from the cell model calculations at $\eta_r=\eta_g=0.028$ from which we find approximate ``effective" surface potentials $\phi_g=5.5$ and $\phi_r=1.2$. When we use these parameters to determine $\Phi(d)$ shown in Fig.  \ref{fig:twobody}(e), we see a much stronger ``short-distance" attraction than in Fig.  \ref{fig:twobody}(a), which extends to distances on the order of a few microns, beyond which the long-range repulsion sets in. The attraction is again accompanied by the alignment of the induced dipoles, shown in the inset of Fig.  \ref{fig:twobody}(e). In fact, at lower $d$ we even see that $Z_r<0$, although $Z_r+Z_g>0$ at contact, so that the resulting dumbbell is positively charged. We emphasize that although the effective potential of Fig. \ref{fig:twobody}(e) resembles in some sense the DLVO potential, the situation here is rather different. The short-distance attraction here is not due to the Van der Waals attraction which is induced by electronic polarization, but occurs here purely due to \emph{ionic} electrostatics. This is also reflected by the micron-range of the present attraction, which is much longer than that of a typical van der Waals attraction.

Finally, we calculated the effective interaction potential between a red-green dumbbell and a red or green particle in Fig. \ref{fig:dumbbelint}, with $\phi_r=1.2$ on the ``red'' part of the dumbbell and $\phi_g=5.5$ on a ``green'' surface. We always let the spherical colloid approach the dumbbell along its symmetry axis, such that we can still exploit cylindrical symmetry. When we determine $\Phi$ as a function of the surface-to-surface distance $\Delta d$ we see in Fig. \ref{fig:dumbbelint}(a) that the ``rg-g" repulsions between a red-green dumbbell and a green colloid approaching the green part of the dumbbell is little different from the ``g-g'' (green-green) repulsion. In contrast, the ``gr-r'' repulsion between a green-red dumbbell and a red particle is rather different: Fig. \ref{fig:dumbbelint}(b) shows that red particles are more strongly repelled from the red parts of the dumbbell than the ``r-r'' repulsion between two single red particles. The ``r-g'' attraction between red and green particles shown in Fig. \ref{fig:dumbbelint}(c) is however similar to the red-green and red attraction ``rg-r'', with the only difference that the net free energy gain is larger at contact for  the interaction between single red and green particles. The ``gr-g'' repulsion at large $d$ between a green-red dumbbell and a green particle is, however, stronger than the r-g and rg-r repulsion at large $d$ and the energy barrier is higher. The calculation as presented in this section and the previous one will now be related to the experiments of section \ref{sec:experimental}.

\section{Speculation on alternating string and cluster formation in binary systems}
Using the results from the binary cell model of section \ref{sec:bincell} and the two-body calculations of section \ref{sec:twobody}, we can now give an explanation for the observation of the alternating strings and clusters from section \ref{sec:form}. In Fig. \ref{fig:twobody}(b), we have seen that a short-distance attraction is produced by matching the charges of the two colloids at $d\rightarrow\infty$ to the charges as obtained from the binary cell model of Fig. \ref{fig:charges}, emphasizing the important role of many-body effects in this system. The energy barrier is, however, quite high, $\Delta E\sim 80\ k_BT$. For an attempt frequency per particle of $\nu=1 \ \mathrm{s}^{-1}$,  the production rate of a red-green dumbbell per particle present in the system is $\nu\exp(-{\beta\Delta E})\sim10^{-35}\ \mathrm{s}^{-1}$, which is too low compared to the experimental time scales. 

\begin{figure}[t]
\includegraphics[width=0.51\textwidth]{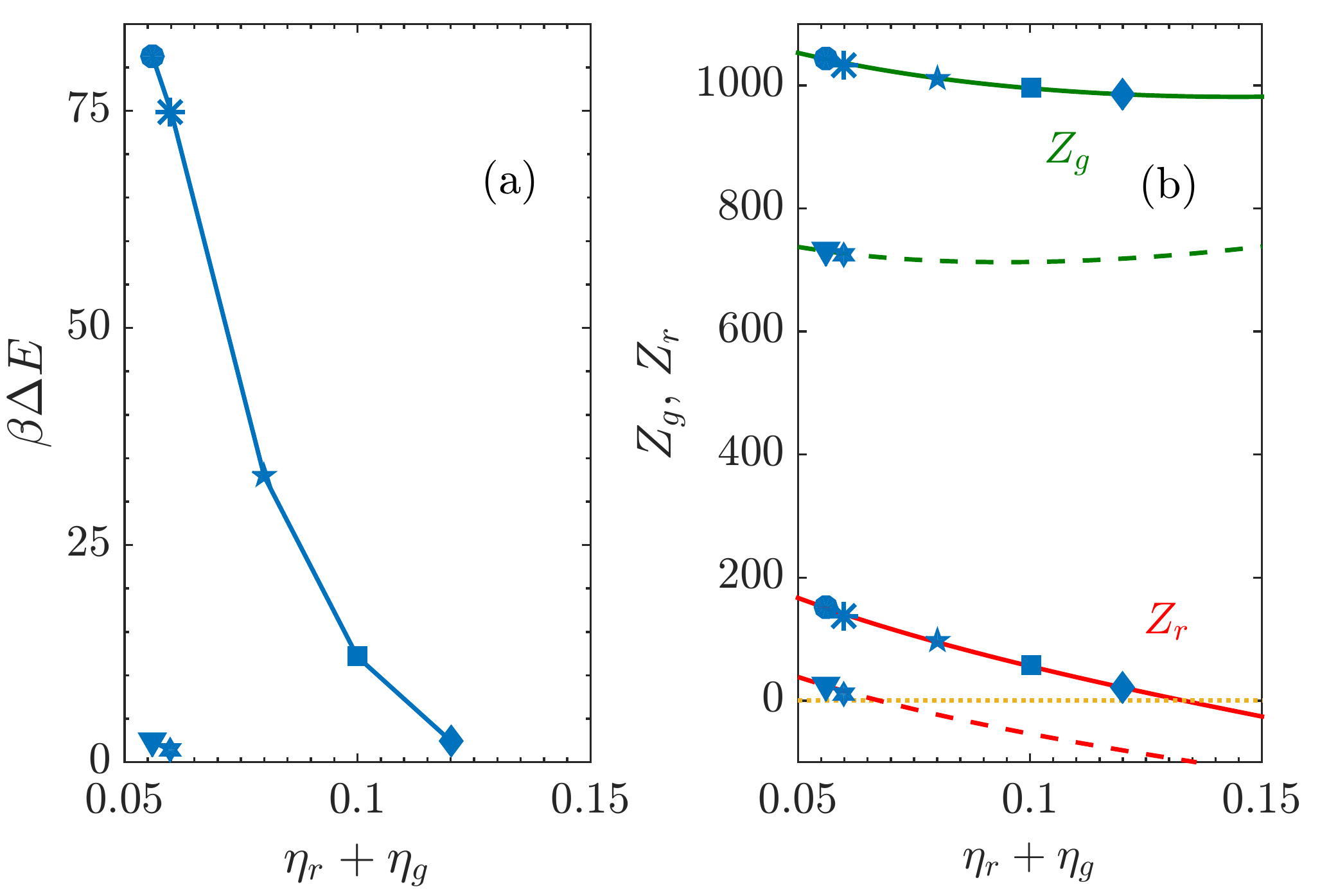}
\caption{(a) Energy barrier that separates a regime of short-distance attraction and long-range repulsion for various state points using the bulk charges from the binary cell model of Fig. \ref{fig:charges} with $\eta_r=\eta_g$. The lines are used to guide the eye. The upper line is for $\kappa^{-1}=6\ \mu\mathrm{m}$, while the lower one {\mycolor(two points at the bottom)} is for a lower ionic strength, $\kappa^{-1}=10\ \mu\mathrm{m}$, and in (b) we plot the corresponding charges of red ($Z_r$) and green ($Z_g$) particles. The full lines are for $\kappa^{-1}=6\ \mu\mathrm{m}$ and the dashed lines are for $\kappa^{-1}=10\ \mu\mathrm{m}$, and the symbols match the state points of (a).}
\label{fig:barrier}
\end{figure}

Similar to the reasoning why the short-distance attraction can be enhanced by including discharging due to many-body effects, we hypothesize that the energy barrier can  be sufficiently lowered if we allow for variations in the \emph{local} volume fraction. To demonstrate this, we use the charges as obtained from the binary cell model of Fig. \ref{fig:charges} at $\kappa^{-1}=6\ \mu\mathrm{m}$ and $\kappa^{-1}=10\ \mu\mathrm{m}$ as function of $\eta_r+\eta_g$ with $\eta_r=\eta_g$. Using effective surface potentials to match the charges of the cell model to the charges within the two-body approximation at $d\rightarrow\infty$, we determined $\Delta E$ in Fig. \ref{fig:barrier}(a) for various state points. Furthermore, we plot the corresponding (bulk) charges in Fig. \ref{fig:barrier}(b). We see that (local) variations in the volume fraction can lower $\Delta E$ through particle discharging, and in particular the energy barrier is lowest when the red particle is close to being charge neutral. If the total (local) volume fraction is twice as large, we find $\Delta E\sim 10\ k_BT$, with a corresponding production rate per particle of a red-green dumbbell $\sim 10^{-5} \ \mathrm{s}^{-1}$, which means that a single green-red dumbbell is produced within $\sim 6$ hours per particle, close to the experimentally observed time scales. The production rate is even larger at a higher local volume fraction, showing the sensitivity of the energy barrier to the local colloid density. At a slighly lower ionic strength, we see that the energy barrier is even only a few $k_B T$, see Fig. \ref{fig:barrier}, and that charge inversion of the red particle occurs above $\eta_r+\eta_g\sim0.6$.

Because of a short-distance attraction which is accompanied by the alignment of the net dipole moments of the two particles, we speculate that we find strings instead of (spherically symmetric) clusters. Moreover, van der Waals forces are only important on length scales of a few nanometers, which is much smaller than the range of attractions set by $\kappa^{-1}$ we observe here. Furthermore, it is important that the alignment of the dipole moments occurs at a larger $d$ than the one for which charge inversion of the red particle takes place, see section \ref{sec:bincell}. This implies that the energy barrier for a particle with an aligned dipole moment is lower, and hence, it is more probable for particles approaching an existing dumbbell along its symmetry axis. Other possibilities, say an approach of a particle not exactly along the symmetry axis of an existing dumbbell, can have an energy barrier that is only a few $k_B T$ higher, and may therefore also occur, rationalizing why the bond angle within a string is not always 180 degrees. An ion sink such as water reduces the energy barrier even more (compare bottom symbols and top symbols in Fig. \ref{fig:barrier}(a)), explaining the more extended strings in Fig. \ref{CHfig:str}(b) compared to Fig. \ref{CHfig:str}(a).

Why (extended) clusters are found near the oil-water interface can also be understood from our calculations. Close to the interface, we observed a larger local density, and hence from Fig. \ref{fig:barrier}, we find that $\Delta E$ is even lower than it would be in bulk (filled circle). The energy barriers between various angles on which a particle can approach a green-red dumbbells may differ only by a few $k_BT$ and hence all of them can occur within an appreciable timescale, explaining the clusters. We hypothesize that there is still a small energy barrier, because the clusters were linearly extended and not spherically symmetric. Although it is not clear how much the ionic strength is reduced due to the presence of the water phase, our results do suggest that a larger Debye length will indeed promote the formation of the clusters through charge regulation, in line with Ref. \cite{Nakamura:2015}.
Finally, the relatively favourable rg-r and gr-g interactions support the fact that the strings and clusters are alternating, although more calculations are needed to study the height of the barrier that separates the long-range repulsions from the shorter ranged attractions, for example, also within a true three-dimensional geometry, where the cylindrical symmetry cannot be exploited.

\section{Open questions on dumbbell formation in one-component systems}
In previous experiments \cite{vdLinden:2013} we compressed dispersions of sterically stabilized charged PMMA particles in CHB by centrifugation, and subsequently followed the system over several months. We found that a significant fraction of the particles had formed small clusters during the centrifugation step. The fraction of clustered particles decreased in time due to spontaneous dissociation of the clusters, as the ionic strength in the dispersion increased due to decomposition of the solvent CHB. We found similar behaviour (cluster formation and dissociation) when the particles were pressed together by an electric field \cite{vdLinden:2013}.

\begin{figure}[t]
\centering
\includegraphics[width=0.5\textwidth]{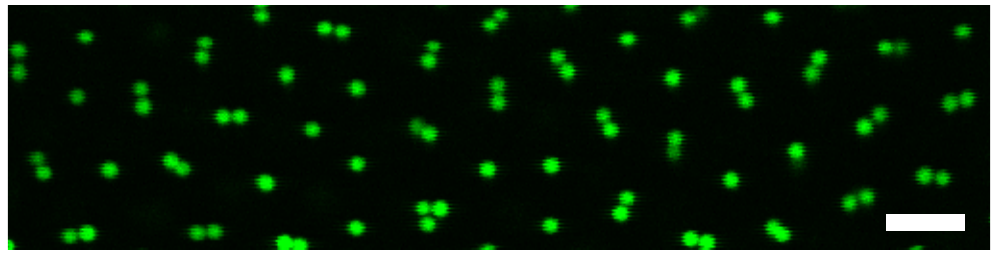}
\caption{Clusters (mostly dumbbells) formed in a suspension of locked PMMA particles (radius $a=1.12$~$\mu$m) in CHB at a distance  0.3 mm from the CHB-water interface, two days after sample preparation. \mycolor{The scale bar indicates 10 $\mu$m.}}
\label{fig:dumbbellformation}
\end{figure}

\begin{figure}[t]
\centering
\includegraphics[width=0.45\textwidth]{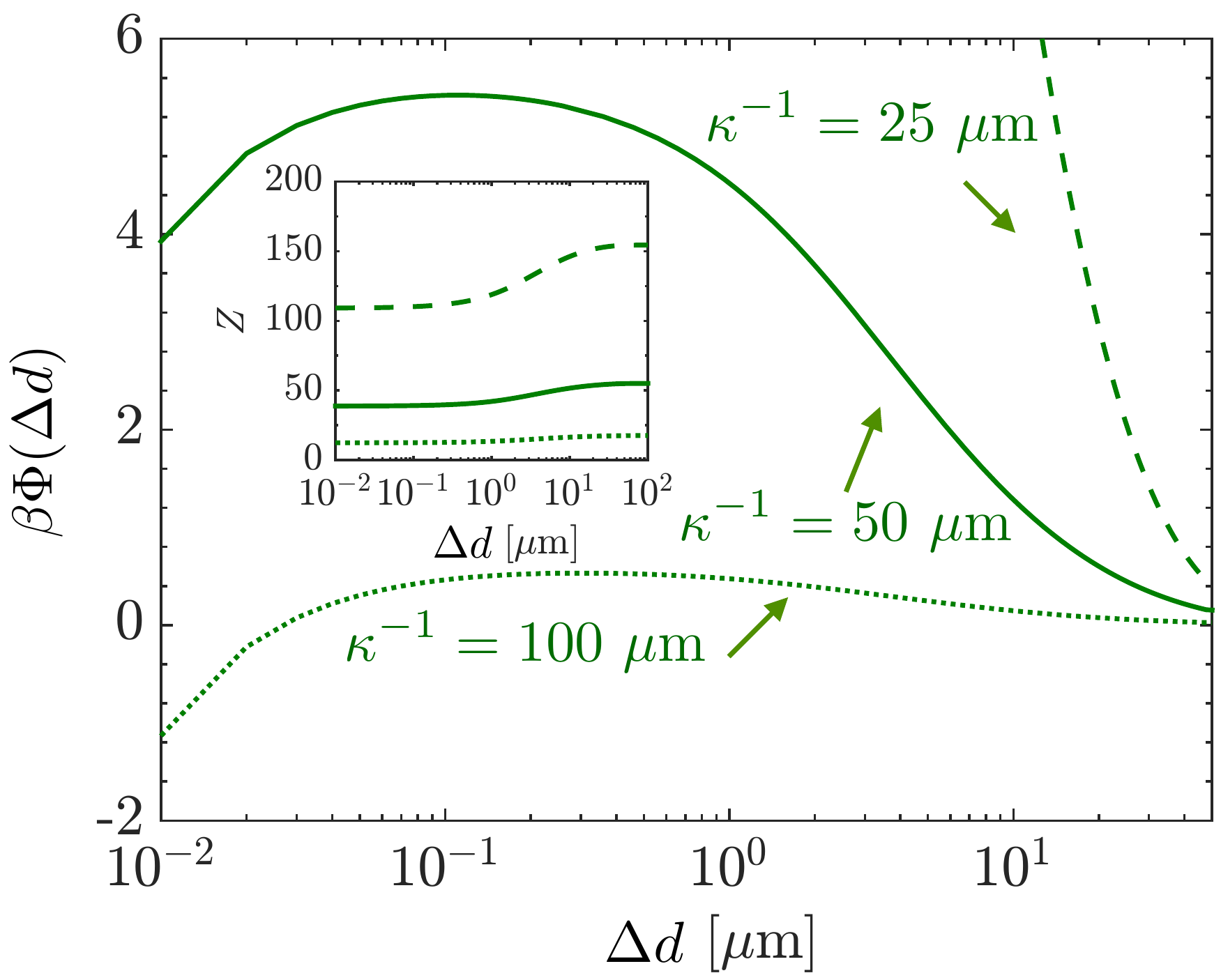}
\caption{Colloid-colloid effective pair potentials $\Phi$ as function of surface-to-surface distance $\Delta d$ for two colloids of radius $a=1.12\ \mu\mathrm{m}$ at various Debye screening length $\kappa^{-1}$ . A sphere-sphere van der Waals attraction has been added with Hamaker constant $A_H=0.2\ k_BT$, while for $\Delta d<10\ \mathrm{nm}$ there is a hard-sphere repulsion to model the effect of steric layer with a thickness of $5$ nm. For each $\kappa^{-1}$, we used an effective surface potential that gives a $Z$ that coincides at $\Delta d\rightarrow\infty$ with the result of a spherical-cell model where we used a surface potential $\phi_0=4.6$ for all state points.}
\label{fig:ggpotential}
\end{figure}

Interestingly, in a one-component dispersion of green particles with radius $a=1.12\ \mu\mathrm{m}$ and surface potential $\phi_0=4.6$ (system 7 in Ref. \cite{Linden:2015}), we also observed the formation of dumbbells similarly to the system of Ref. \cite{vdLinden:2013}, provided that the system was brought into contact with a water phase, see Fig. \ref{fig:dumbbellformation}, where a few green-green dumbbells are shown. This  observation cannot be explained by the two-body approximation of the previous section, because we found that the r-r and g-g interaction between a pair of the same species is always repulsive.
There are now a few possible scenarios that we will consider. The first one is that adding a water phase adjacent to the oil-phase generates a Donnan potential \cite{Leunissen2:2007} across the oil-water interface due to ion partitioning, and hence a local electric field near the interface is generated that may induce dipole moments on the particles that in turn generate an attraction when these are aligned. However, we do not observe any alignment of the dumbbells in a preferred direction, so this seems an unlikely mechanism.
The second one is that the water phase acts as an ion sink that increases $\kappa^{-1}$ in the oil, leading to discharging of the green particles. The repulsions then become weaker and this may render the van der Waals attraction to be more important, such that the observed clusters could  be a consequence of microphase separation within conventional DLVO theory  \cite{Groenewold:2001, Groenewold:2004}. To test this hypothesis, we estimate the strength of the van der Waals interaction $\Phi_\text{VdW}$ between two equal-sized spheres within Hamaker-de Boer theory, which describes the pair potential (neglecting retardation and screening) as
\begin{equation}
\beta\Phi_\text{VdW}(d)=-\frac{A_H}{3}\left[\frac{a^2}{d^2-4a^2}+\frac{a^2}{d^2}+\frac{1}{2}\ln\left(1-\frac{4a^2}{d^2}\right)\right].\label{eq:vdwspheres}
\end{equation}
As is well known, this expression fails at very short distances $d\approx 2a$ where the atom-atom Born repulsion plays an important role, and for large distances where retardation becomes important due to the finite speed of light. In general, a cutoff for the surface-to-surface distance of $0.16$ nm \cite{Israelachvili} is used to estimate the maximum attraction that is possible due to van der Waals.
We approximate the Hamaker constant $A_H$ by the static contribution within Lifschitz theory \cite{Lifschitz:1954}, such that
\begin{equation}
\beta A_H=\frac{3}{4}\left(\frac{{{\mycolor{\epsilon_s}}}-\epsilon_c}{{\mycolor{\epsilon_s}}+\epsilon_c}\right)^2,
\end{equation}
which gives the lower bound $\beta A_H=0.2$, since we neglected the (positive) contribution from the summation over all frequencies . On this basis, we then estimate that the attraction on closest approach $\Phi_\text{VdW}\sim-100\ k_BT$. However, the PHSA-PMMA particles that we use have a steric layer that reduces the van der Waals attraction even further (closest approach $\sim 10 \ \mathrm{nm}$), which makes the van der Waals attraction on contact even weaker, namely $\sim -2 \ k_BT$. To test whether this attraction is strong enough to overcome the electrostratic repulsions, we used for all state points a spherical cell model with a surface potential $\phi_0=4.6$ \cite{Linden:2015}. From it we determined effective surface potentials to be used within the two-body approximation with the condition that the two-body $Z(d)$ coincides with the cell model result for $d\rightarrow\infty$. In Fig. \ref{fig:ggpotential} we show the resulting pair potentials $\Phi(d)$ with the van der Waals attraction of Eq. \eqref{eq:vdwspheres} added and with $Z(d)$ in the inset. We see that the screening length in oil must be {\mycolor increased} to $\kappa^{-1}\sim50\ \mu\mathrm{m}$ for a (metastable) bound state to occur with an energy barrier of $\sim 5 \ k_BT$. For a stable bound state the ionic strength must be reduced even further through the uptake of ions by the water phase. In Fig.  \ref{fig:ggpotential} we show for example the resulting $\Phi$ for $\kappa^{-1}=100\ \mu\mathrm{m}$, showing a stable bound state separated from the bulk with an energy barrier of $\sim 1\ k_BT$. 

Note that the values of $\kappa^{-1}$ are quite large, but can still give rise to a sufficient charge $Z$ for the colloids within a two-body approximation. One could think that such long Debye lengths are rather unrealistic in a true many-body suspension, where the spheres would be completely discharged at such a state point. However, we stress that $\kappa^{-1}$ is the Debye length of the \emph{reservoir}. We have only performed many-body corrections to the surface potential, but not to the Debye length. In principle, this can be taken into account within the cell model by defining $\bar{\kappa}^2=\kappa^2\cosh(\phi_D)$. This is also the quantity one needs to use in Debye-H\"uckel theory to correct for non-linear effects involving charge renormalization \cite{Trizac:2002} and $\bar{\kappa}^{-1}$ can therefore be seen as an effective screening length due to double layer overlaps. Indeed, for a reservoir screening length of $\kappa^{-1}=100\ \mu\mathrm{m}$ at $\eta=0.02$ and $\phi_0=4.6$, we find using the cell model $\bar{\kappa}^{-1}=13 \ \mu\mathrm{m}$, a numerical value that is maybe more intuitive in a true many-body system regarding the values of $Z$ involved. Therefore, the calculation as performed in Fig. \ref{fig:ggpotential} is maybe a poor estimate for the range of the interactions. However, because the charges are not so sensitive to $\rho_s$ in the dilute limit, Fig. \ref{fig:ggpotential} is a good estimate for the height of the energy barrier , which is more of interest at this stage.


Finally, Fig. \ref{fig:dumbbellformation} shows that the dumbbells seem to order in a plastic crystal, at a volume fraction $\eta\sim 0.02$. We test this hypothesis by calculating the OCP coupling parameter $\Gamma$ for this system within a cell model for $\kappa^{-1}=6$ $\mu$m. The procedure for such a calculation can be found in Refs. \cite{Boon:2015, Everts:2016}, where it is shown that $\Gamma>106$ is a reliable freezing criterion. We found that decreasing $\rho_s$ at fixed $\eta_g$ reduces $\Gamma$, so the presence of water will actually impede crystallization. However, we remark that within such a calculation we assume that each lattice point is occupied by a single particle, while dumbbells have a larger charge and hence their existence tends to increase $\Gamma$. In fact, we find that our results are rather sensitive to the precise values of the parameters such as $\kappa^{-1}$, the local colloid density and the charges of the constituent particles. Hence, we could not precisely assess whether the system is in a crystalline state or not within the theory. However, considering the reported surface potential for these particles $\phi_0=4.6$ \cite{Linden:2015} we see that $\Gamma$ varies between $\Gamma=260$ at $\kappa^{-1}=6\ \mu\mathrm{m}$, to $\Gamma=137$ at $\kappa^{-1}=10\ \mu\mathrm{m}$ and $\Gamma= 70$ at $\kappa^{-1}=15\ \mu\mathrm{m}$, For the metastable bound state at $\kappa^{-1}=50\ \mu\mathrm{m}$ in Fig. \ref{fig:ggpotential}, however, we find $\Gamma=3.1$, and hence according to the theory the system is actually far from being crystalline for repulsions that can be overcome by a small van der Waals attraction of $\sim$-2 $k_BT$. However, given the sensitivity to $\kappa^{-1}$ there is certainly a possibility that the system is close to a crystallization transition. Therefore, another alternative explanation is that the dumbbell formation can be seen as the formation of a Wigner crystal with a double occupancy of particles for some of the lattice sites, which shows a striking resemblance to the one discussed in Ref. \cite{Mladek:2008} for a much shorter-ranged repulsion. 

The precise mechanism for the formation of these dumbbells is thus still an open question. {\mycolor However, we are tempted to favour the multiple-occupancy crystal over the formation of a (meta-)stable dimer state by a residual van der Waals attraction. Namely, in the latter case one needs (i) finetuning of the system parameters to enter a regime in which clusters would form and (ii) one would also expect higher order clusters. For a multiple occupancy crystal, however, no finetuning is needed as it can be driven purely by repulsions. Secondly, higher order occupancy of lattice sites is limited in this case by the hard-core and electrostatic repulsion of the particles. It is therefore conceivable that many-body effects not only select the lattice spacing, but also the mean occupancy number of the lattice sites.  It is interesting to investigate the dumbbell formation systematically in experiments and theory, which we will both leave for future work. }

\section{Discussion and conclusion}
We investigated a system of size- and charge-asymmetric colloids with constant-potential boundary conditions. We found that within the two-body approximation it is possible to have a net attraction between the two like-charged particles that induces the formation of alternating strings if (i) the charge of one of the two particles is low enough and (ii) the induced dipole moments are aligned. We studied many-body effects in the effective pair interaction by using effective surface potentials that stem from a spherical-cell model, providing short-distance attractions through charge regulation. Moreover, another additional feature is the formation of aligned induced dipoles that may favor alternating strings over compact clusters. Such strings are not to be expected from standard DLVO theory where the charge is assumed to be spatially constant over the colloidal surface.  Moreover, we have shown that enhanced cluster formation is expected to occur upon lowering the ionic strength of the binary suspensions, which enlarges the regime of composition and overall packing fraction in which the colloids are found to be oppositely charged. It would be interesting to investigate the induced-dipole interactions in a true many-body system, although this is a challenging problem in general. Perhaps an engaging way to tackle this is through the Car-Parinello like simulation techniques first proposed by Fushiki \cite{Fushiki:1992} and L\"owen \emph{et al.} \cite{Lowen:1992, Lowen:1993}, which offers the possibility to include the effects of deformed ionic screening clouds and the fact that the effective potential between $N$ colloids depends on the many-body configuration of colloids through charge regulation. We hypothesize that these effects are of utmost importance for the formation of alternating strings and clusters in binary suspensions of charged colloids at the large screening lengths in these systems. This is left for future work. Finally, we have speculated that{\mycolor dimers and trimers found in single-component dispersions of PHSA-PMMA particles can be seen as the result of a multiple-occupancy Wigner crystal, and also this hypothesis can be tested with the methods of Refs. \cite{Fushiki:1992, Lowen:1992, Lowen:1993}.} \\

\begin{acknowledgements}

{\mycolor J.C.E. and M.N.v.d.L. contributed equally to this work, J.C.E. carried the calculations, M.N.v.d.L. performed the experiments, A.v.B. and R.v.R. initiated the project and supervised the experiments and theory, respectively, and all authors contributed to the writing of the manuscript.} We acknowledge fruitful discussions with Tara Drwenski and Sela Samin. {\mycolor We thank Johan Stiefelhagen and G\"ul\c{s}en Heessels-G\"urbo\u{g}a for particle synthesis}. Furthermore, we acknowledge financial support of a Netherlands Organisation for Scientific Research (NWO) VICI grant {\mycolor and an NWO Toptalent grant} funded by the Dutch Ministry of Education, Culture and Science (OCW). This work is part of the D-ITP consortium, a program of the Netherlands Organisation for Scientific Research (NWO) funded by the Dutch Ministry of Education, Culture and Science (OCW). {\mycolor Finally, we also acknowledge excellent suggestions from an anonymous referee on the double occupancy of the Wigner crystal.}
\end{acknowledgements}
\bibliographystyle{apsrev4-1} 
\bibliography{literature1.bib} 

\begin{thebibliography}{64}%
\makeatletter
\providecommand \@ifxundefined [1]{%
 \@ifx{#1\undefined}
}%
\providecommand \@ifnum [1]{%
 \ifnum #1\expandafter \@firstoftwo
 \else \expandafter \@secondoftwo
 \fi
}%
\providecommand \@ifx [1]{%
 \ifx #1\expandafter \@firstoftwo
 \else \expandafter \@secondoftwo
 \fi
}%
\providecommand \natexlab [1]{#1}%
\providecommand \enquote  [1]{``#1''}%
\providecommand \bibnamefont  [1]{#1}%
\providecommand \bibfnamefont [1]{#1}%
\providecommand \citenamefont [1]{#1}%
\providecommand \href@noop [0]{\@secondoftwo}%
\providecommand \href [0]{\begingroup \@sanitize@url \@href}%
\providecommand \@href[1]{\@@startlink{#1}\@@href}%
\providecommand \@@href[1]{\endgroup#1\@@endlink}%
\providecommand \@sanitize@url [0]{\catcode `\\12\catcode `\$12\catcode
  `\&12\catcode `\#12\catcode `\^12\catcode `\_12\catcode `\%12\relax}%
\providecommand \@@startlink[1]{}%
\providecommand \@@endlink[0]{}%
\providecommand \url  [0]{\begingroup\@sanitize@url \@url }%
\providecommand \@url [1]{\endgroup\@href {#1}{\urlprefix }}%
\providecommand \urlprefix  [0]{URL }%
\providecommand \Eprint [0]{\href }%
\providecommand \doibase [0]{http://dx.doi.org/}%
\providecommand \selectlanguage [0]{\@gobble}%
\providecommand \bibinfo  [0]{\@secondoftwo}%
\providecommand \bibfield  [0]{\@secondoftwo}%
\providecommand \translation [1]{[#1]}%
\providecommand \BibitemOpen [0]{}%
\providecommand \bibitemStop [0]{}%
\providecommand \bibitemNoStop [0]{.\EOS\space}%
\providecommand \EOS [0]{\spacefactor3000\relax}%
\providecommand \BibitemShut  [1]{\csname bibitem#1\endcsname}%
\let\auto@bib@innerbib\@empty
\bibitem [{\citenamefont {Jones}(2002)}]{Jones}%
  \BibitemOpen
  \bibfield  {author} {\bibinfo {author} {\bibfnamefont {R.~A.~L.}\
  \bibnamefont {Jones}},\ }in\ \href@noop {} {\emph {\bibinfo {booktitle} {Soft
  Condensed Matter}}}\ (\bibinfo  {publisher} {OUP Oxford},\ \bibinfo {year}
  {2002})\BibitemShut {NoStop}%
\bibitem [{\citenamefont {Alberts}\ \emph {et~al.}(2007)\citenamefont
  {Alberts}, \citenamefont {Johnson}, \citenamefont {Lewis}, \citenamefont
  {Raff}, \citenamefont {Roberts},\ and\ \citenamefont {Walter}}]{Alberts}%
  \BibitemOpen
  \bibfield  {author} {\bibinfo {author} {\bibfnamefont {B.}~\bibnamefont
  {Alberts}}, \bibinfo {author} {\bibfnamefont {A.}~\bibnamefont {Johnson}},
  \bibinfo {author} {\bibfnamefont {J.}~\bibnamefont {Lewis}}, \bibinfo
  {author} {\bibfnamefont {M.}~\bibnamefont {Raff}}, \bibinfo {author}
  {\bibfnamefont {K.}~\bibnamefont {Roberts}}, \ and\ \bibinfo {author}
  {\bibfnamefont {P.}~\bibnamefont {Walter}},\ }in\ \href@noop {} {\emph
  {\bibinfo {booktitle} {Molecular Biology of the Cell (5th edition)}}}\
  (\bibinfo  {publisher} {Garland Science},\ \bibinfo {year}
  {2007})\BibitemShut {NoStop}%
\bibitem [{\citenamefont {Chaikin}\ and\ \citenamefont
  {Lubensky}(2000)}]{Lubensky}%
  \BibitemOpen
  \bibfield  {author} {\bibinfo {author} {\bibfnamefont {P.~M.}\ \bibnamefont
  {Chaikin}}\ and\ \bibinfo {author} {\bibfnamefont {T.~C.}\ \bibnamefont
  {Lubensky}},\ }in\ \href@noop {} {\emph {\bibinfo {booktitle} {Principles of
  condensed matter physics}}}\ (\bibinfo  {publisher} {Cambridge University
  Press},\ \bibinfo {year} {2000})\BibitemShut {NoStop}%
\bibitem [{\citenamefont {Vroege}\ and\ \citenamefont
  {Lekkerkerker}(1992)}]{Vroege:1992}%
  \BibitemOpen
  \bibfield  {author} {\bibinfo {author} {\bibfnamefont {G.~J.}\ \bibnamefont
  {Vroege}}\ and\ \bibinfo {author} {\bibfnamefont {H.~N.~W.}\ \bibnamefont
  {Lekkerkerker}},\ }\href {http://stacks.iop.org/0034-4885/55/i=8/a=003}
  {\bibfield  {journal} {\bibinfo  {journal} {Rep. Prog. Phys.}\ }\textbf
  {\bibinfo {volume} {55}},\ \bibinfo {pages} {1241} (\bibinfo {year}
  {1992})}\BibitemShut {NoStop}%
\bibitem [{\citenamefont {Timmermans}(1961)}]{Timmermans:1961}%
  \BibitemOpen
  \bibfield  {author} {\bibinfo {author} {\bibfnamefont {J.}~\bibnamefont
  {Timmermans}},\ }\href {\doibase
  http://dx.doi.org/10.1016/0022-3697(61)90076-2} {\bibfield  {journal}
  {\bibinfo  {journal} {Journal of Physics and Chemistry of Solids}\ }\textbf
  {\bibinfo {volume} {18}},\ \bibinfo {pages} {1 } (\bibinfo {year}
  {1961})}\BibitemShut {NoStop}%
\bibitem [{\citenamefont {Liu}\ \emph {et~al.}(2013)\citenamefont {Liu},
  \citenamefont {Besseling}, \citenamefont {Hermes}, \citenamefont
  {Demir\"ors}, \citenamefont {Imhog},\ and\ \citenamefont {van
  Blaaderen}}]{Liu:2014}%
  \BibitemOpen
  \bibfield  {author} {\bibinfo {author} {\bibfnamefont {B.}~\bibnamefont
  {Liu}}, \bibinfo {author} {\bibfnamefont {T.}~\bibnamefont {Besseling}},
  \bibinfo {author} {\bibfnamefont {M.}~\bibnamefont {Hermes}}, \bibinfo
  {author} {\bibfnamefont {A.}~\bibnamefont {Demir\"ors}}, \bibinfo {author}
  {\bibfnamefont {A.}~\bibnamefont {Imhog}}, \ and\ \bibinfo {author}
  {\bibfnamefont {A.}~\bibnamefont {van Blaaderen}},\ }\href
  {http://www.nature.com/ncomms/2014/140121/ncomms4092/full/ncomms4092.html}
  {\bibfield  {journal} {\bibinfo  {journal} {Nat. Comm.}\ }\textbf {\bibinfo
  {volume} {5}} (\bibinfo {year} {2013})}\BibitemShut {NoStop}%
\bibitem [{\citenamefont {Shechtman}\ \emph {et~al.}(1984)\citenamefont
  {Shechtman}, \citenamefont {Blech}, \citenamefont {Gratias},\ and\
  \citenamefont {Cahn}}]{Schechtman:1984}%
  \BibitemOpen
  \bibfield  {author} {\bibinfo {author} {\bibfnamefont {D.}~\bibnamefont
  {Shechtman}}, \bibinfo {author} {\bibfnamefont {I.}~\bibnamefont {Blech}},
  \bibinfo {author} {\bibfnamefont {D.}~\bibnamefont {Gratias}}, \ and\
  \bibinfo {author} {\bibfnamefont {J.~W.}\ \bibnamefont {Cahn}},\ }\href
  {\doibase 10.1103/PhysRevLett.53.1951} {\bibfield  {journal} {\bibinfo
  {journal} {Phys. Rev. Lett.}\ }\textbf {\bibinfo {volume} {53}},\ \bibinfo
  {pages} {1951} (\bibinfo {year} {1984})}\BibitemShut {NoStop}%
\bibitem [{\citenamefont {Talapin}\ \emph {et~al.}(2009)\citenamefont
  {Talapin}, \citenamefont {Shevchenko}, \citenamefont {Bodnarchuk},
  \citenamefont {Ye}, \citenamefont {Chen},\ and\ \citenamefont
  {Murray}}]{Talapin:2009}%
  \BibitemOpen
  \bibfield  {author} {\bibinfo {author} {\bibfnamefont {D.}~\bibnamefont
  {Talapin}}, \bibinfo {author} {\bibfnamefont {E.}~\bibnamefont {Shevchenko}},
  \bibinfo {author} {\bibfnamefont {M.}~\bibnamefont {Bodnarchuk}}, \bibinfo
  {author} {\bibfnamefont {X.}~\bibnamefont {Ye}}, \bibinfo {author}
  {\bibfnamefont {J.}~\bibnamefont {Chen}}, \ and\ \bibinfo {author}
  {\bibfnamefont {C.}~\bibnamefont {Murray}},\ }\href {\doibase
  10.1038/nature08439} {\bibfield  {journal} {\bibinfo  {journal} {Nature}\
  }\textbf {\bibinfo {volume} {461}},\ \bibinfo {pages} {964} (\bibinfo {year}
  {2009})}\BibitemShut {NoStop}%
\bibitem [{\citenamefont {Genc}\ and\ \citenamefont {Derin}(2014)}]{Genc:2014}%
  \BibitemOpen
  \bibfield  {author} {\bibinfo {author} {\bibfnamefont {S.}~\bibnamefont
  {Genc}}\ and\ \bibinfo {author} {\bibfnamefont {B.}~\bibnamefont {Derin}},\
  }\href {\doibase http://dx.doi.org/10.1016/j.coche.2013.12.006} {\bibfield
  {journal} {\bibinfo  {journal} {Curr. Op. Chem. Eng.}\ }\textbf {\bibinfo
  {volume} {3}},\ \bibinfo {pages} {118 } (\bibinfo {year} {2014})}\BibitemShut
  {NoStop}%
\bibitem [{\citenamefont {Watson}(1953)}]{Watson:1953}%
  \BibitemOpen
  \bibfield  {author} {\bibinfo {author} {\bibfnamefont {J.~D.}\ \bibnamefont
  {Watson}},\ }\href
  {http://www.nature.com/physics/looking-back/crick/Crick_Watson.pdf}
  {\bibfield  {journal} {\bibinfo  {journal} {Nature}\ }\textbf {\bibinfo
  {volume} {171}},\ \bibinfo {pages} {737—738} (\bibinfo {year}
  {1953})}\BibitemShut {NoStop}%
\bibitem [{\citenamefont {Dill}\ and\ \citenamefont
  {MacCallum}(2012)}]{Dill:2012}%
  \BibitemOpen
  \bibfield  {author} {\bibinfo {author} {\bibfnamefont {K.~A.}\ \bibnamefont
  {Dill}}\ and\ \bibinfo {author} {\bibfnamefont {J.~L.}\ \bibnamefont
  {MacCallum}},\ }\href {\doibase 10.1126/science.1219021} {\bibfield
  {journal} {\bibinfo  {journal} {Science}\ }\textbf {\bibinfo {volume}
  {338}},\ \bibinfo {pages} {1042} (\bibinfo {year} {2012})}\BibitemShut
  {NoStop}%
\bibitem [{\citenamefont {Leibler}(1980)}]{Leibler:1980}%
  \BibitemOpen
  \bibfield  {author} {\bibinfo {author} {\bibfnamefont {L.}~\bibnamefont
  {Leibler}},\ }\href {\doibase 10.1021/ma60078a047} {\bibfield  {journal}
  {\bibinfo  {journal} {Macromolecules}\ }\textbf {\bibinfo {volume} {13}},\
  \bibinfo {pages} {1602} (\bibinfo {year} {1980})}\BibitemShut {NoStop}%
\bibitem [{\citenamefont {Russel}\ \emph {et~al.}(1992)\citenamefont {Russel},
  \citenamefont {Saville},\ and\ \citenamefont {Schowalter}}]{Russel}%
  \BibitemOpen
  \bibfield  {author} {\bibinfo {author} {\bibfnamefont {W.~B.}\ \bibnamefont
  {Russel}}, \bibinfo {author} {\bibfnamefont {D.~A.}\ \bibnamefont {Saville}},
  \ and\ \bibinfo {author} {\bibfnamefont {W.~R.}\ \bibnamefont {Schowalter}},\
  }in\ \href@noop {} {\emph {\bibinfo {booktitle} {Colloidal Dispersions}}}\
  (\bibinfo  {publisher} {Cambridge University Press},\ \bibinfo {year}
  {1992})\BibitemShut {NoStop}%
\bibitem [{\citenamefont {Yethiraj}(2007)}]{Anand:2007}%
  \BibitemOpen
  \bibfield  {author} {\bibinfo {author} {\bibfnamefont {A.}~\bibnamefont
  {Yethiraj}},\ }\href {\doibase 10.1039/B704251P} {\bibfield  {journal}
  {\bibinfo  {journal} {Soft Matter}\ }\textbf {\bibinfo {volume} {3}},\
  \bibinfo {pages} {1099} (\bibinfo {year} {2007})}\BibitemShut {NoStop}%
\bibitem [{\citenamefont {Ninham}\ and\ \citenamefont
  {Parsegian}(1971)}]{Ninham:1971}%
  \BibitemOpen
  \bibfield  {author} {\bibinfo {author} {\bibfnamefont {B.~W.}\ \bibnamefont
  {Ninham}}\ and\ \bibinfo {author} {\bibfnamefont {V.}~\bibnamefont
  {Parsegian}},\ }\href {\doibase 10.1016/0022-5193(71)90019-1} {\bibfield
  {journal} {\bibinfo  {journal} {J. Theor. Biol.}\ }\textbf {\bibinfo {volume}
  {31}},\ \bibinfo {pages} {405 } (\bibinfo {year} {1971})}\BibitemShut
  {NoStop}%
\bibitem [{\citenamefont {Yethiraj}\ and\ \citenamefont {van
  Blaaderen}(2003)}]{Blaaderen:2003}%
  \BibitemOpen
  \bibfield  {author} {\bibinfo {author} {\bibfnamefont {A.}~\bibnamefont
  {Yethiraj}}\ and\ \bibinfo {author} {\bibfnamefont {A.}~\bibnamefont {van
  Blaaderen}},\ }\href {\doibase 10.1038/nature01328} {\bibfield  {journal}
  {\bibinfo  {journal} {Nature}\ }\textbf {\bibinfo {volume} {421}},\ \bibinfo
  {pages} {513} (\bibinfo {year} {2003})}\BibitemShut {NoStop}%
\bibitem [{\citenamefont {van Roij}\ and\ \citenamefont
  {Hansen}(1997)}]{Roij:1997}%
  \BibitemOpen
  \bibfield  {author} {\bibinfo {author} {\bibfnamefont {R.}~\bibnamefont {van
  Roij}}\ and\ \bibinfo {author} {\bibfnamefont {J.-P.}\ \bibnamefont
  {Hansen}},\ }\href {\doibase 10.1103/PhysRevLett.79.3082} {\bibfield
  {journal} {\bibinfo  {journal} {Phys. Rev. Lett.}\ }\textbf {\bibinfo
  {volume} {79}},\ \bibinfo {pages} {3082} (\bibinfo {year}
  {1997})}\BibitemShut {NoStop}%
\bibitem [{\citenamefont {Zoetekouw}(2006)}]{Zoetekouw}%
  \BibitemOpen
  \bibfield  {author} {\bibinfo {author} {\bibfnamefont {B.}~\bibnamefont
  {Zoetekouw}},\ }\emph {\bibinfo {title} {Phase behavior of charged colloids
  many-body effects, charge renormalization and charge regulation}},\ \href
  {http://dspace.library.uu.nl/bitstream/handle/1874/12536/index.htm?sequence=2}
  {Ph.D. thesis},\ \bibinfo  {school} {Utrecht University} (\bibinfo {year}
  {2006})\BibitemShut {NoStop}%
\bibitem [{\citenamefont {L\"owen}\ \emph {et~al.}(1993)\citenamefont
  {L\"owen}, \citenamefont {Hansen},\ and\ \citenamefont
  {Madden}}]{Lowen2:1993}%
  \BibitemOpen
  \bibfield  {author} {\bibinfo {author} {\bibfnamefont {H.}~\bibnamefont
  {L\"owen}}, \bibinfo {author} {\bibfnamefont {J.}~\bibnamefont {Hansen}}, \
  and\ \bibinfo {author} {\bibfnamefont {P.~A.}\ \bibnamefont {Madden}},\
  }\href
  {http://scitation.aip.org/content/aip/journal/jcp/98/4/10.1063/1.464099}
  {\bibfield  {journal} {\bibinfo  {journal} {J. Chem. Phys.}\ }\textbf
  {\bibinfo {volume} {98}},\ \bibinfo {pages} {3275} (\bibinfo {year}
  {1993})}\BibitemShut {NoStop}%
\bibitem [{\citenamefont {Warren}(2000)}]{Warren:2000}%
  \BibitemOpen
  \bibfield  {author} {\bibinfo {author} {\bibfnamefont {P.~B.}\ \bibnamefont
  {Warren}},\ }\href
  {http://scitation.aip.org/content/aip/journal/jcp/112/10/10.1063/1.481024}
  {\bibfield  {journal} {\bibinfo  {journal} {J. Chem. Phys.}\ }\textbf
  {\bibinfo {volume} {112}},\ \bibinfo {pages} {4683} (\bibinfo {year}
  {2000})}\BibitemShut {NoStop}%
\bibitem [{\citenamefont {Denton}(2000)}]{Denton:2005}%
  \BibitemOpen
  \bibfield  {author} {\bibinfo {author} {\bibfnamefont {A.~R.}\ \bibnamefont
  {Denton}},\ }\href {\doibase 10.1103/PhysRevE.62.3855} {\bibfield  {journal}
  {\bibinfo  {journal} {Phys. Rev. E}\ }\textbf {\bibinfo {volume} {62}},\
  \bibinfo {pages} {3855} (\bibinfo {year} {2000})}\BibitemShut {NoStop}%
\bibitem [{\citenamefont {Zoetekouw}\ and\ \citenamefont {van
  Roij}(2006)}]{Zoetekouw:2006}%
  \BibitemOpen
  \bibfield  {author} {\bibinfo {author} {\bibfnamefont {B.}~\bibnamefont
  {Zoetekouw}}\ and\ \bibinfo {author} {\bibfnamefont {R.}~\bibnamefont {van
  Roij}},\ }\href {\doibase 10.1103/PhysRevLett.97.258302} {\bibfield
  {journal} {\bibinfo  {journal} {Phys. Rev. Lett.}\ }\textbf {\bibinfo
  {volume} {97}},\ \bibinfo {pages} {258302} (\bibinfo {year}
  {2006})}\BibitemShut {NoStop}%
\bibitem [{\citenamefont {Israelachvili}(2011)}]{Israelachvili}%
  \BibitemOpen
  \bibfield  {author} {\bibinfo {author} {\bibfnamefont {J.~N.}\ \bibnamefont
  {Israelachvili}},\ }in\ \href@noop {} {\emph {\bibinfo {booktitle}
  {Intermolecular and Surface Forces (Third Edition)}}}\ (\bibinfo  {publisher}
  {Elsevier},\ \bibinfo {year} {2011})\BibitemShut {NoStop}%
\bibitem [{\citenamefont {Derjaguin}\ and\ \citenamefont
  {Landau}(1941)}]{Derjaguin:1948}%
  \BibitemOpen
  \bibfield  {author} {\bibinfo {author} {\bibfnamefont {B.}~\bibnamefont
  {Derjaguin}}\ and\ \bibinfo {author} {\bibfnamefont {L.}~\bibnamefont
  {Landau}},\ }\href {\doibase 10.1016/0079-6816(93)90013-L} {\bibfield
  {journal} {\bibinfo  {journal} {Acta Physicochim. URSS}\ }\textbf {\bibinfo
  {volume} {14}},\ \bibinfo {pages} {633} (\bibinfo {year} {1941})}\BibitemShut
  {NoStop}%
\bibitem [{\citenamefont {Verwey}\ and\ \citenamefont
  {Overbeek}(1948)}]{VerweyOverbeek}%
  \BibitemOpen
  \bibfield  {author} {\bibinfo {author} {\bibfnamefont {E.~J.~W.}\
  \bibnamefont {Verwey}}\ and\ \bibinfo {author} {\bibfnamefont {J.~T.~G.}\
  \bibnamefont {Overbeek}},\ }in\ \href@noop {} {\emph {\bibinfo {booktitle}
  {Theory of the Stability of Lyophobic Colloids}}}\ (\bibinfo  {publisher}
  {Elsevier, New York},\ \bibinfo {year} {1948})\BibitemShut {NoStop}%
\bibitem [{\citenamefont {Nakamura}\ \emph {et~al.}(2015)\citenamefont
  {Nakamura}, \citenamefont {Okachi}, \citenamefont {Toyotama}, \citenamefont
  {Okuzono},\ and\ \citenamefont {Yamanaka}}]{Nakamura:2015}%
  \BibitemOpen
  \bibfield  {author} {\bibinfo {author} {\bibfnamefont {Y.}~\bibnamefont
  {Nakamura}}, \bibinfo {author} {\bibfnamefont {M.}~\bibnamefont {Okachi}},
  \bibinfo {author} {\bibfnamefont {A.}~\bibnamefont {Toyotama}}, \bibinfo
  {author} {\bibfnamefont {T.}~\bibnamefont {Okuzono}}, \ and\ \bibinfo
  {author} {\bibfnamefont {J.}~\bibnamefont {Yamanaka}},\ }\href {\doibase
  10.1021/acs.langmuir.5b02778} {\bibfield  {journal} {\bibinfo  {journal}
  {Langmuir}\ }\textbf {\bibinfo {volume} {31}},\ \bibinfo {pages} {13303}
  (\bibinfo {year} {2015})}\BibitemShut {NoStop}%
\bibitem [{\citenamefont {Hoffmann}\ \emph {et~al.}(2004)\citenamefont
  {Hoffmann}, \citenamefont {Likos},\ and\ \citenamefont
  {Hansen}}]{Hoffman:2004}%
  \BibitemOpen
  \bibfield  {author} {\bibinfo {author} {\bibfnamefont {N.}~\bibnamefont
  {Hoffmann}}, \bibinfo {author} {\bibfnamefont {C.~N.}\ \bibnamefont {Likos}},
  \ and\ \bibinfo {author} {\bibfnamefont {J.-P.}\ \bibnamefont {Hansen}},\
  }\href {\doibase 10.1080/00268970410001695688} {\bibfield  {journal}
  {\bibinfo  {journal} {Mol. Phys.}\ }\textbf {\bibinfo {volume} {102}},\
  \bibinfo {pages} {857} (\bibinfo {year} {2004})}\BibitemShut {NoStop}%
\bibitem [{\citenamefont {Ramirez}\ and\ \citenamefont
  {Kjellander}(2006)}]{Ramirez:2006}%
  \BibitemOpen
  \bibfield  {author} {\bibinfo {author} {\bibfnamefont {R.}~\bibnamefont
  {Ramirez}}\ and\ \bibinfo {author} {\bibfnamefont {R.}~\bibnamefont
  {Kjellander}},\ }\href
  {http://scitation.aip.org/content/aip/journal/jcp/125/14/10.1063/1.2355486}
  {\bibfield  {journal} {\bibinfo  {journal} {J. Chem. Phys.}\ }\textbf
  {\bibinfo {volume} {125}},\ \bibinfo {eid} {144110} (\bibinfo {year}
  {2006})}\BibitemShut {NoStop}%
\bibitem [{\citenamefont {Boon}\ \emph {et~al.}(2010)\citenamefont {Boon},
  \citenamefont {Gallardo}, \citenamefont {Zheng}, \citenamefont {Eggen},
  \citenamefont {Dijkstra},\ and\ \citenamefont {van Roij}}]{Boon:2010}%
  \BibitemOpen
  \bibfield  {author} {\bibinfo {author} {\bibfnamefont {N.}~\bibnamefont
  {Boon}}, \bibinfo {author} {\bibfnamefont {E.~C.}\ \bibnamefont {Gallardo}},
  \bibinfo {author} {\bibfnamefont {S.}~\bibnamefont {Zheng}}, \bibinfo
  {author} {\bibfnamefont {E.}~\bibnamefont {Eggen}}, \bibinfo {author}
  {\bibfnamefont {M.}~\bibnamefont {Dijkstra}}, \ and\ \bibinfo {author}
  {\bibfnamefont {R.}~\bibnamefont {van Roij}},\ }\href
  {http://stacks.iop.org/0953-8984/22/i=10/a=104104} {\bibfield  {journal}
  {\bibinfo  {journal} {J. Phys.: Cond. Matt.}\ }\textbf {\bibinfo {volume}
  {22}},\ \bibinfo {pages} {104104} (\bibinfo {year} {2010})}\BibitemShut
  {NoStop}%
\bibitem [{\citenamefont {de~Graaf}\ \emph {et~al.}(2012)\citenamefont
  {de~Graaf}, \citenamefont {Boon}, \citenamefont {Dijkstra},\ and\
  \citenamefont {van Roij}}]{Graaf:2012}%
  \BibitemOpen
  \bibfield  {author} {\bibinfo {author} {\bibfnamefont {J.}~\bibnamefont
  {de~Graaf}}, \bibinfo {author} {\bibfnamefont {N.}~\bibnamefont {Boon}},
  \bibinfo {author} {\bibfnamefont {M.}~\bibnamefont {Dijkstra}}, \ and\
  \bibinfo {author} {\bibfnamefont {R.}~\bibnamefont {van Roij}},\ }\href
  {\doibase 10.1063/1.4751482} {\bibfield  {journal} {\bibinfo  {journal} {J.
  Chem. Phys.}\ }\textbf {\bibinfo {volume} {137}},\ \bibinfo {eid} {104910}
  (\bibinfo {year} {2012}),\ 10.1063/1.4751482}\BibitemShut {NoStop}%
\bibitem [{\citenamefont {Smallenburg}\ \emph {et~al.}(2012)\citenamefont
  {Smallenburg}, \citenamefont {Vutukuri}, \citenamefont {Imhof}, \citenamefont
  {van Blaaderen},\ and\ \citenamefont {Dijkstra}}]{Smallenburg:2012}%
  \BibitemOpen
  \bibfield  {author} {\bibinfo {author} {\bibfnamefont {F.}~\bibnamefont
  {Smallenburg}}, \bibinfo {author} {\bibfnamefont {H.~R.}\ \bibnamefont
  {Vutukuri}}, \bibinfo {author} {\bibfnamefont {A.}~\bibnamefont {Imhof}},
  \bibinfo {author} {\bibfnamefont {A.}~\bibnamefont {van Blaaderen}}, \ and\
  \bibinfo {author} {\bibfnamefont {M.}~\bibnamefont {Dijkstra}},\ }\href
  {http://stacks.iop.org/0953-8984/24/i=46/a=464113} {\bibfield  {journal}
  {\bibinfo  {journal} {J.Phys.: Cond. Matt.}\ }\textbf {\bibinfo {volume}
  {24}},\ \bibinfo {pages} {464113} (\bibinfo {year} {2012})}\BibitemShut
  {NoStop}%
\bibitem [{\citenamefont {Leunissen}\ \emph {et~al.}(2005)\citenamefont
  {Leunissen}, \citenamefont {Christova}, \citenamefont {Hynninen},
  \citenamefont {Royall}, \citenamefont {Campbell}, \citenamefont {Imhof},
  \citenamefont {Dijkstra}, \citenamefont {van Roij},\ and\ \citenamefont {van
  Blaaderen}}]{Leunissen:2005}%
  \BibitemOpen
  \bibfield  {author} {\bibinfo {author} {\bibfnamefont {M.~E.}\ \bibnamefont
  {Leunissen}}, \bibinfo {author} {\bibfnamefont {C.~G.}\ \bibnamefont
  {Christova}}, \bibinfo {author} {\bibfnamefont {A.-P.}\ \bibnamefont
  {Hynninen}}, \bibinfo {author} {\bibfnamefont {C.~P.}\ \bibnamefont
  {Royall}}, \bibinfo {author} {\bibfnamefont {A.~I.}\ \bibnamefont
  {Campbell}}, \bibinfo {author} {\bibfnamefont {A.}~\bibnamefont {Imhof}},
  \bibinfo {author} {\bibfnamefont {M.}~\bibnamefont {Dijkstra}}, \bibinfo
  {author} {\bibfnamefont {R.}~\bibnamefont {van Roij}}, \ and\ \bibinfo
  {author} {\bibfnamefont {A.}~\bibnamefont {van Blaaderen}},\ }\href {\doibase
  10.1038/nature03946} {\bibfield  {journal} {\bibinfo  {journal} {Nature}\
  }\textbf {\bibinfo {volume} {437}},\ \bibinfo {pages} {235} (\bibinfo {year}
  {2005})}\BibitemShut {NoStop}%
\bibitem [{\citenamefont {Vissers}\ \emph
  {et~al.}(2011{\natexlab{a}})\citenamefont {Vissers}, \citenamefont {Wysocki},
  \citenamefont {Rex}, \citenamefont {Lowen}, \citenamefont {Royall},
  \citenamefont {Imhof},\ and\ \citenamefont {van Blaaderen}}]{Vissers:2011}%
  \BibitemOpen
  \bibfield  {author} {\bibinfo {author} {\bibfnamefont {T.}~\bibnamefont
  {Vissers}}, \bibinfo {author} {\bibfnamefont {A.}~\bibnamefont {Wysocki}},
  \bibinfo {author} {\bibfnamefont {M.}~\bibnamefont {Rex}}, \bibinfo {author}
  {\bibfnamefont {H.}~\bibnamefont {Lowen}}, \bibinfo {author} {\bibfnamefont
  {C.~P.}\ \bibnamefont {Royall}}, \bibinfo {author} {\bibfnamefont
  {A.}~\bibnamefont {Imhof}}, \ and\ \bibinfo {author} {\bibfnamefont
  {A.}~\bibnamefont {van Blaaderen}},\ }\href {\doibase 10.1039/C0SM01343A}
  {\bibfield  {journal} {\bibinfo  {journal} {Soft Matter}\ }\textbf {\bibinfo
  {volume} {7}},\ \bibinfo {pages} {2352} (\bibinfo {year}
  {2011}{\natexlab{a}})}\BibitemShut {NoStop}%
\bibitem [{\citenamefont {Sear}\ and\ \citenamefont
  {Gelbart}(1999)}]{Sear:1999}%
  \BibitemOpen
  \bibfield  {author} {\bibinfo {author} {\bibfnamefont {R.~P.}\ \bibnamefont
  {Sear}}\ and\ \bibinfo {author} {\bibfnamefont {W.~M.}\ \bibnamefont
  {Gelbart}},\ }\href {\doibase 10.1063/1.478338} {\bibfield  {journal}
  {\bibinfo  {journal} {J. Chem. Phys.}\ }\textbf {\bibinfo {volume} {110}},\
  \bibinfo {pages} {4582} (\bibinfo {year} {1999})}\BibitemShut {NoStop}%
\bibitem [{\citenamefont {Stradner}\ \emph {et~al.}(2004)\citenamefont
  {Stradner}, \citenamefont {Sedgwick}, \citenamefont {Cardinaux},
  \citenamefont {Poon}, \citenamefont {Egelhaaf},\ and\ \citenamefont
  {Schurtenberger}}]{stradner:2004}%
  \BibitemOpen
  \bibfield  {author} {\bibinfo {author} {\bibfnamefont {A.}~\bibnamefont
  {Stradner}}, \bibinfo {author} {\bibfnamefont {H.}~\bibnamefont {Sedgwick}},
  \bibinfo {author} {\bibfnamefont {F.}~\bibnamefont {Cardinaux}}, \bibinfo
  {author} {\bibfnamefont {W.~C.}\ \bibnamefont {Poon}}, \bibinfo {author}
  {\bibfnamefont {S.~U.}\ \bibnamefont {Egelhaaf}}, \ and\ \bibinfo {author}
  {\bibfnamefont {P.}~\bibnamefont {Schurtenberger}},\ }\href {\doibase
  10.1038/nature03109} {\bibfield  {journal} {\bibinfo  {journal} {Nature}\
  }\textbf {\bibinfo {volume} {432}},\ \bibinfo {pages} {492} (\bibinfo {year}
  {2004})}\BibitemShut {NoStop}%
\bibitem [{\citenamefont {Imperio}\ and\ \citenamefont
  {Reatto}(2004)}]{Imperio:2004}%
  \BibitemOpen
  \bibfield  {author} {\bibinfo {author} {\bibfnamefont {A.}~\bibnamefont
  {Imperio}}\ and\ \bibinfo {author} {\bibfnamefont {L.}~\bibnamefont
  {Reatto}},\ }\href {http://stacks.iop.org/0953-8984/16/i=38/a=001} {\bibfield
   {journal} {\bibinfo  {journal} {J. Phys.: Cond. Matt.}\ }\textbf {\bibinfo
  {volume} {16}},\ \bibinfo {pages} {S3769} (\bibinfo {year}
  {2004})}\BibitemShut {NoStop}%
\bibitem [{\citenamefont {Archer}\ and\ \citenamefont
  {Wilding}(2007)}]{Archer:2007}%
  \BibitemOpen
  \bibfield  {author} {\bibinfo {author} {\bibfnamefont {A.~J.}\ \bibnamefont
  {Archer}}\ and\ \bibinfo {author} {\bibfnamefont {N.~B.}\ \bibnamefont
  {Wilding}},\ }\href {\doibase 10.1103/PhysRevE.76.031501} {\bibfield
  {journal} {\bibinfo  {journal} {Phys. Rev. E}\ }\textbf {\bibinfo {volume}
  {76}},\ \bibinfo {pages} {031501} (\bibinfo {year} {2007})}\BibitemShut
  {NoStop}%
\bibitem [{\citenamefont {Groenewold}\ and\ \citenamefont
  {Kegel}(2001)}]{Groenewold:2001}%
  \BibitemOpen
  \bibfield  {author} {\bibinfo {author} {\bibfnamefont {J.}~\bibnamefont
  {Groenewold}}\ and\ \bibinfo {author} {\bibfnamefont {W.~K.}\ \bibnamefont
  {Kegel}},\ }\href {\doibase 10.1021/jp011646w} {\bibfield  {journal}
  {\bibinfo  {journal} {J. Phys. Chem. B}\ }\textbf {\bibinfo {volume} {105}},\
  \bibinfo {pages} {11702} (\bibinfo {year} {2001})}\BibitemShut {NoStop}%
\bibitem [{\citenamefont {Groenewold}\ and\ \citenamefont
  {Kegel}(2004)}]{Groenewold:2004}%
  \BibitemOpen
  \bibfield  {author} {\bibinfo {author} {\bibfnamefont {J.}~\bibnamefont
  {Groenewold}}\ and\ \bibinfo {author} {\bibfnamefont {W.~K.}\ \bibnamefont
  {Kegel}},\ }\href@noop {} {\bibfield  {journal} {\bibinfo  {journal} {J.
  Phys.: Cond. Matt.}\ }\textbf {\bibinfo {volume} {16}},\ \bibinfo {pages}
  {S4877} (\bibinfo {year} {2004})}\BibitemShut {NoStop}%
\bibitem [{\citenamefont {Barros}\ and\ \citenamefont
  {Luijten}(2014)}]{Luijten:2014}%
  \BibitemOpen
  \bibfield  {author} {\bibinfo {author} {\bibfnamefont {K.}~\bibnamefont
  {Barros}}\ and\ \bibinfo {author} {\bibfnamefont {E.}~\bibnamefont
  {Luijten}},\ }\href {\doibase 10.1103/PhysRevLett.113.017801} {\bibfield
  {journal} {\bibinfo  {journal} {Phys. Rev. Lett.}\ }\textbf {\bibinfo
  {volume} {113}},\ \bibinfo {pages} {017801} (\bibinfo {year}
  {2014})}\BibitemShut {NoStop}%
\bibitem [{\citenamefont {Vissers}\ \emph
  {et~al.}(2011{\natexlab{b}})\citenamefont {Vissers}, \citenamefont {Imhof},
  \citenamefont {Carrique}, \citenamefont {Ángel V.~Delgado},\ and\
  \citenamefont {van Blaaderen}}]{Vissers2:2011}%
  \BibitemOpen
  \bibfield  {author} {\bibinfo {author} {\bibfnamefont {T.}~\bibnamefont
  {Vissers}}, \bibinfo {author} {\bibfnamefont {A.}~\bibnamefont {Imhof}},
  \bibinfo {author} {\bibfnamefont {F.}~\bibnamefont {Carrique}}, \bibinfo
  {author} {\bibnamefont {Ángel V.~Delgado}}, \ and\ \bibinfo {author}
  {\bibfnamefont {A.}~\bibnamefont {van Blaaderen}},\ }\href {\doibase
  http://dx.doi.org/10.1016/j.jcis.2011.04.113} {\bibfield  {journal} {\bibinfo
   {journal} {J. Coll. Int. Sci.}\ }\textbf {\bibinfo {volume} {361}},\
  \bibinfo {pages} {443 } (\bibinfo {year} {2011}{\natexlab{b}})}\BibitemShut
  {NoStop}%
\bibitem [{\citenamefont {van~der Linden}\ \emph {et~al.}(2015)\citenamefont
  {van~der Linden}, \citenamefont {Stiefelhagen}, \citenamefont
  {Heessels-Gürboğa}, \citenamefont {van~der Hoeven}, \citenamefont {Elbers},
  \citenamefont {Dijkstra},\ and\ \citenamefont {van Blaaderen}}]{Linden:2015}%
  \BibitemOpen
  \bibfield  {author} {\bibinfo {author} {\bibfnamefont {M.~N.}\ \bibnamefont
  {van~der Linden}}, \bibinfo {author} {\bibfnamefont {J.~C.~P.}\ \bibnamefont
  {Stiefelhagen}}, \bibinfo {author} {\bibfnamefont {G.}~\bibnamefont
  {Heessels-Gürboğa}}, \bibinfo {author} {\bibfnamefont {J.~E.~S.}\
  \bibnamefont {van~der Hoeven}}, \bibinfo {author} {\bibfnamefont {N.~A.}\
  \bibnamefont {Elbers}}, \bibinfo {author} {\bibfnamefont {M.}~\bibnamefont
  {Dijkstra}}, \ and\ \bibinfo {author} {\bibfnamefont {A.}~\bibnamefont {van
  Blaaderen}},\ }\href {\doibase 10.1021/la503665e} {\bibfield  {journal}
  {\bibinfo  {journal} {Langmuir}\ }\textbf {\bibinfo {volume} {31}},\ \bibinfo
  {pages} {65} (\bibinfo {year} {2015})}\BibitemShut {NoStop}%
\bibitem [{\citenamefont {Morrison}(1993)}]{Morrison:1993}%
  \BibitemOpen
  \bibfield  {author} {\bibinfo {author} {\bibfnamefont {I.~D.}\ \bibnamefont
  {Morrison}},\ }\href {\doibase
  http://dx.doi.org/10.1016/0927-7757(93)80026-B} {\bibfield  {journal}
  {\bibinfo  {journal} {Colloids and Surfaces A}\ }\textbf {\bibinfo {volume}
  {71}},\ \bibinfo {pages} {1 } (\bibinfo {year} {1993})}\BibitemShut {NoStop}%
\bibitem [{\citenamefont {van~der Linden}\ \emph {et~al.}(2013)\citenamefont
  {van~der Linden}, \citenamefont {El~Masri}, \citenamefont {Dijkstra},\ and\
  \citenamefont {van Blaaderen}}]{vdLinden:2013}%
  \BibitemOpen
  \bibfield  {author} {\bibinfo {author} {\bibfnamefont {M.~N.}\ \bibnamefont
  {van~der Linden}}, \bibinfo {author} {\bibfnamefont {D.}~\bibnamefont
  {El~Masri}}, \bibinfo {author} {\bibfnamefont {M.}~\bibnamefont {Dijkstra}},
  \ and\ \bibinfo {author} {\bibfnamefont {A.}~\bibnamefont {van Blaaderen}},\
  }\href {\doibase 10.1039/C3SM51752G} {\bibfield  {journal} {\bibinfo
  {journal} {Soft Matter}\ }\textbf {\bibinfo {volume} {9}},\ \bibinfo {pages}
  {11618} (\bibinfo {year} {2013})}\BibitemShut {NoStop}%
\bibitem [{sup()}]{supp}%
  \BibitemOpen
  \href@noop {} {\emph {\bibinfo {title} {Supplementary
  Information}}}\BibitemShut {NoStop}%
\bibitem [{\citenamefont {Leunissen}\ \emph
  {et~al.}(2007{\natexlab{a}})\citenamefont {Leunissen}, \citenamefont {van
  Blaaderen}, \citenamefont {Hollingsworth}, \citenamefont {Sullivan},\ and\
  \citenamefont {Chaikin}}]{Leunissen:2007}%
  \BibitemOpen
  \bibfield  {author} {\bibinfo {author} {\bibfnamefont {M.~E.}\ \bibnamefont
  {Leunissen}}, \bibinfo {author} {\bibfnamefont {A.}~\bibnamefont {van
  Blaaderen}}, \bibinfo {author} {\bibfnamefont {A.~D.}\ \bibnamefont
  {Hollingsworth}}, \bibinfo {author} {\bibfnamefont {M.~T.}\ \bibnamefont
  {Sullivan}}, \ and\ \bibinfo {author} {\bibfnamefont {P.~M.}\ \bibnamefont
  {Chaikin}},\ }\href {\doibase 10.1073/pnas.0610589104} {\bibfield  {journal}
  {\bibinfo  {journal} {PNAS}\ }\textbf {\bibinfo {volume} {104}},\ \bibinfo
  {pages} {2585} (\bibinfo {year} {2007}{\natexlab{a}})}\BibitemShut {NoStop}%
\bibitem [{\citenamefont {Smallenburg}\ \emph {et~al.}(2011)\citenamefont
  {Smallenburg}, \citenamefont {Boon}, \citenamefont {Kater}, \citenamefont
  {Dijkstra},\ and\ \citenamefont {van Roij}}]{Smallenburg:2011}%
  \BibitemOpen
  \bibfield  {author} {\bibinfo {author} {\bibfnamefont {F.}~\bibnamefont
  {Smallenburg}}, \bibinfo {author} {\bibfnamefont {N.}~\bibnamefont {Boon}},
  \bibinfo {author} {\bibfnamefont {M.}~\bibnamefont {Kater}}, \bibinfo
  {author} {\bibfnamefont {M.}~\bibnamefont {Dijkstra}}, \ and\ \bibinfo
  {author} {\bibfnamefont {R.}~\bibnamefont {van Roij}},\ }\href
  {http://scitation.aip.org/content/aip/journal/jcp/134/7/10.1063/1.3555627}
  {\bibfield  {journal} {\bibinfo  {journal} {J. Chem. Phys.}\ }\textbf
  {\bibinfo {volume} {134}},\ \bibinfo {eid} {074505} (\bibinfo {year}
  {2011})}\BibitemShut {NoStop}%
\bibitem [{\citenamefont {Alexander}\ \emph {et~al.}(1984)\citenamefont
  {Alexander}, \citenamefont {Chaikin}, \citenamefont {Grant}, \citenamefont
  {Morales}, \citenamefont {Pincus},\ and\ \citenamefont
  {Hone}}]{Alexander:1984}%
  \BibitemOpen
  \bibfield  {author} {\bibinfo {author} {\bibfnamefont {S.}~\bibnamefont
  {Alexander}}, \bibinfo {author} {\bibfnamefont {P.~M.}\ \bibnamefont
  {Chaikin}}, \bibinfo {author} {\bibfnamefont {P.}~\bibnamefont {Grant}},
  \bibinfo {author} {\bibfnamefont {G.~J.}\ \bibnamefont {Morales}}, \bibinfo
  {author} {\bibfnamefont {P.}~\bibnamefont {Pincus}}, \ and\ \bibinfo {author}
  {\bibfnamefont {D.}~\bibnamefont {Hone}},\ }\href {\doibase 10.1063/1.446600}
  {\bibfield  {journal} {\bibinfo  {journal} {J. Chem. Phys.}\ }\textbf
  {\bibinfo {volume} {80}},\ \bibinfo {pages} {5776} (\bibinfo {year}
  {1984})}\BibitemShut {NoStop}%
\bibitem [{\citenamefont {Tamashiro}\ \emph {et~al.}(1998)\citenamefont
  {Tamashiro}, \citenamefont {Levin},\ and\ \citenamefont
  {Barbosa}}]{Tamashiro:1998}%
  \BibitemOpen
  \bibfield  {author} {\bibinfo {author} {\bibfnamefont {M.}~\bibnamefont
  {Tamashiro}}, \bibinfo {author} {\bibfnamefont {Y.}~\bibnamefont {Levin}}, \
  and\ \bibinfo {author} {\bibfnamefont {M.}~\bibnamefont {Barbosa}},\ }\href
  {\doibase 10.1007/s100510050192} {\bibfield  {journal} {\bibinfo  {journal}
  {Eur. Phys. J. B}\ }\textbf {\bibinfo {volume} {1}},\ \bibinfo {pages} {337}
  (\bibinfo {year} {1998})}\BibitemShut {NoStop}%
\bibitem [{\citenamefont {Trizac}\ \emph {et~al.}(2002)\citenamefont {Trizac},
  \citenamefont {Bocquet},\ and\ \citenamefont {Aubouy}}]{Trizac:2002}%
  \BibitemOpen
  \bibfield  {author} {\bibinfo {author} {\bibfnamefont {E.}~\bibnamefont
  {Trizac}}, \bibinfo {author} {\bibfnamefont {L.}~\bibnamefont {Bocquet}}, \
  and\ \bibinfo {author} {\bibfnamefont {M.}~\bibnamefont {Aubouy}},\ }\href
  {\doibase 10.1103/PhysRevLett.89.248301} {\bibfield  {journal} {\bibinfo
  {journal} {Phys. Rev. Lett.}\ }\textbf {\bibinfo {volume} {89}},\ \bibinfo
  {pages} {248301} (\bibinfo {year} {2002})}\BibitemShut {NoStop}%
\bibitem [{\citenamefont {Eggen}\ and\ \citenamefont {van
  Roij}(2009)}]{Eggen:2009}%
  \BibitemOpen
  \bibfield  {author} {\bibinfo {author} {\bibfnamefont {E.}~\bibnamefont
  {Eggen}}\ and\ \bibinfo {author} {\bibfnamefont {R.}~\bibnamefont {van
  Roij}},\ }\href {\doibase 10.1103/PhysRevE.80.041402} {\bibfield  {journal}
  {\bibinfo  {journal} {Phys. Rev. E}\ }\textbf {\bibinfo {volume} {80}},\
  \bibinfo {pages} {041402} (\bibinfo {year} {2009})}\BibitemShut {NoStop}%
\bibitem [{\citenamefont {Denton}(2010)}]{Denton:2010}%
  \BibitemOpen
  \bibfield  {author} {\bibinfo {author} {\bibfnamefont {A.~R.}\ \bibnamefont
  {Denton}},\ }\href {http://stacks.iop.org/0953-8984/22/i=36/a=364108}
  {\bibfield  {journal} {\bibinfo  {journal} {J. Phys.: Cond. Matt.}\ }\textbf
  {\bibinfo {volume} {22}},\ \bibinfo {pages} {364108} (\bibinfo {year}
  {2010})}\BibitemShut {NoStop}%
\bibitem [{\citenamefont {Valeriani}\ \emph {et~al.}(2010)\citenamefont
  {Valeriani}, \citenamefont {Camp}, \citenamefont {Zwanikken}, \citenamefont
  {van Roij},\ and\ \citenamefont {Dijkstra}}]{Valeriani:2010}%
  \BibitemOpen
  \bibfield  {author} {\bibinfo {author} {\bibfnamefont {C.}~\bibnamefont
  {Valeriani}}, \bibinfo {author} {\bibfnamefont {P.~J.}\ \bibnamefont {Camp}},
  \bibinfo {author} {\bibfnamefont {J.~W.}\ \bibnamefont {Zwanikken}}, \bibinfo
  {author} {\bibfnamefont {R.}~\bibnamefont {van Roij}}, \ and\ \bibinfo
  {author} {\bibfnamefont {M.}~\bibnamefont {Dijkstra}},\ }\href {\doibase
  10.1039/C001577F} {\bibfield  {journal} {\bibinfo  {journal} {Soft Matter}\
  }\textbf {\bibinfo {volume} {6}},\ \bibinfo {pages} {2793} (\bibinfo {year}
  {2010})}\BibitemShut {NoStop}%
\bibitem [{\citenamefont {Torres}\ \emph {et~al.}(2008)\citenamefont {Torres},
  \citenamefont {Téllez},\ and\ \citenamefont {van Roij}}]{Torres:2008}%
  \BibitemOpen
  \bibfield  {author} {\bibinfo {author} {\bibfnamefont {A.}~\bibnamefont
  {Torres}}, \bibinfo {author} {\bibfnamefont {G.}~\bibnamefont {Téllez}}, \
  and\ \bibinfo {author} {\bibfnamefont {R.}~\bibnamefont {van Roij}},\ }\href
  {\doibase 10.1063/1.2907719} {\bibfield  {journal} {\bibinfo  {journal} {J.
  Chem. Phys.}\ }\textbf {\bibinfo {volume} {128}},\ \bibinfo {eid} {154906}
  (\bibinfo {year} {2008})}\BibitemShut {NoStop}%
\bibitem [{\citenamefont {Everts}\ \emph {et~al.}(2016)\citenamefont {Everts},
  \citenamefont {Boon},\ and\ \citenamefont {van Roij}}]{Everts:2016}%
  \BibitemOpen
  \bibfield  {author} {\bibinfo {author} {\bibfnamefont {J.~C.}\ \bibnamefont
  {Everts}}, \bibinfo {author} {\bibfnamefont {N.}~\bibnamefont {Boon}}, \ and\
  \bibinfo {author} {\bibfnamefont {R.}~\bibnamefont {van Roij}},\ }\href
  {\doibase 10.1039/C5CP07943H} {\bibfield  {journal} {\bibinfo  {journal}
  {Phys. Chem. Chem. Phys.}\ }\textbf {\bibinfo {volume} {18}},\ \bibinfo
  {pages} {5211} (\bibinfo {year} {2016})}\BibitemShut {NoStop}%
\bibitem [{\citenamefont {Carnie}\ \emph {et~al.}(1994)\citenamefont {Carnie},
  \citenamefont {Chan},\ and\ \citenamefont {Gunning}}]{Carnie:1994}%
  \BibitemOpen
  \bibfield  {author} {\bibinfo {author} {\bibfnamefont {S.~L.}\ \bibnamefont
  {Carnie}}, \bibinfo {author} {\bibfnamefont {D.~Y.~C.}\ \bibnamefont {Chan}},
  \ and\ \bibinfo {author} {\bibfnamefont {J.~S.}\ \bibnamefont {Gunning}},\
  }\href {\doibase 10.1021/la00021a024} {\bibfield  {journal} {\bibinfo
  {journal} {Langmuir}\ }\textbf {\bibinfo {volume} {10}},\ \bibinfo {pages}
  {2993} (\bibinfo {year} {1994})}\BibitemShut {NoStop}%
\bibitem [{\citenamefont {Stankovich}\ and\ \citenamefont
  {Carnie}(1996)}]{Stankovich:1996}%
  \BibitemOpen
  \bibfield  {author} {\bibinfo {author} {\bibfnamefont {J.}~\bibnamefont
  {Stankovich}}\ and\ \bibinfo {author} {\bibfnamefont {S.~L.}\ \bibnamefont
  {Carnie}},\ }\href {\doibase 10.1021/la950384k} {\bibfield  {journal}
  {\bibinfo  {journal} {Langmuir}\ }\textbf {\bibinfo {volume} {12}},\ \bibinfo
  {pages} {1453} (\bibinfo {year} {1996})}\BibitemShut {NoStop}%
\bibitem [{\citenamefont {Leunissen}\ \emph
  {et~al.}(2007{\natexlab{b}})\citenamefont {Leunissen}, \citenamefont
  {Zwanikken}, \citenamefont {van Roij}, \citenamefont {Chaikin},\ and\
  \citenamefont {van Blaaderen}}]{Leunissen2:2007}%
  \BibitemOpen
  \bibfield  {author} {\bibinfo {author} {\bibfnamefont {M.~E.}\ \bibnamefont
  {Leunissen}}, \bibinfo {author} {\bibfnamefont {J.}~\bibnamefont
  {Zwanikken}}, \bibinfo {author} {\bibfnamefont {R.}~\bibnamefont {van Roij}},
  \bibinfo {author} {\bibfnamefont {P.~M.}\ \bibnamefont {Chaikin}}, \ and\
  \bibinfo {author} {\bibfnamefont {A.}~\bibnamefont {van Blaaderen}},\ }\href
  {\doibase 10.1039/B711300E} {\bibfield  {journal} {\bibinfo  {journal} {Phys.
  Chem. Chem. Phys.}\ }\textbf {\bibinfo {volume} {9}},\ \bibinfo {pages}
  {6405} (\bibinfo {year} {2007}{\natexlab{b}})}\BibitemShut {NoStop}%
\bibitem [{\citenamefont {Lifschitz}()}]{Lifschitz:1954}%
  \BibitemOpen
  \bibfield  {author} {\bibinfo {author} {\bibfnamefont {E.~M.}\ \bibnamefont
  {Lifschitz}},\ }\href@noop {} {\bibfield  {journal} {\bibinfo  {journal} {J.
  Exp. Th. Phys. USSR}\ }\textbf {\bibinfo {volume} {29}},\ \bibinfo {pages}
  {94–110}}\BibitemShut {NoStop}%
\bibitem [{\citenamefont {Kanai}\ \emph {et~al.}(2015)\citenamefont {Kanai},
  \citenamefont {Boon}, \citenamefont {Lu}, \citenamefont {Sloutskin},
  \citenamefont {Schofield}, \citenamefont {Smallenburg}, \citenamefont {van
  Roij}, \citenamefont {Dijkstra},\ and\ \citenamefont {Weitz}}]{Boon:2015}%
  \BibitemOpen
  \bibfield  {author} {\bibinfo {author} {\bibfnamefont {T.}~\bibnamefont
  {Kanai}}, \bibinfo {author} {\bibfnamefont {N.}~\bibnamefont {Boon}},
  \bibinfo {author} {\bibfnamefont {P.~J.}\ \bibnamefont {Lu}}, \bibinfo
  {author} {\bibfnamefont {E.}~\bibnamefont {Sloutskin}}, \bibinfo {author}
  {\bibfnamefont {A.~B.}\ \bibnamefont {Schofield}}, \bibinfo {author}
  {\bibfnamefont {F.}~\bibnamefont {Smallenburg}}, \bibinfo {author}
  {\bibfnamefont {R.}~\bibnamefont {van Roij}}, \bibinfo {author}
  {\bibfnamefont {M.}~\bibnamefont {Dijkstra}}, \ and\ \bibinfo {author}
  {\bibfnamefont {D.~A.}\ \bibnamefont {Weitz}},\ }\href {\doibase
  10.1103/PhysRevE.91.030301} {\bibfield  {journal} {\bibinfo  {journal} {Phys.
  Rev. E}\ }\textbf {\bibinfo {volume} {91}},\ \bibinfo {pages} {030301}
  (\bibinfo {year} {2015})}\BibitemShut {NoStop}%
\bibitem [{\citenamefont {Mladek}\ \emph {et~al.}(2008)\citenamefont {Mladek},
  \citenamefont {Charbonneau}, \citenamefont {Likos}, \citenamefont {Frenkel},\
  and\ \citenamefont {Kahl}}]{Mladek:2008}%
  \BibitemOpen
  \bibfield  {author} {\bibinfo {author} {\bibfnamefont {B.~M.}\ \bibnamefont
  {Mladek}}, \bibinfo {author} {\bibfnamefont {P.}~\bibnamefont {Charbonneau}},
  \bibinfo {author} {\bibfnamefont {C.~N.}\ \bibnamefont {Likos}}, \bibinfo
  {author} {\bibfnamefont {D.}~\bibnamefont {Frenkel}}, \ and\ \bibinfo
  {author} {\bibfnamefont {G.}~\bibnamefont {Kahl}},\ }\href
  {http://stacks.iop.org/0953-8984/20/i=49/a=494245} {\bibfield  {journal}
  {\bibinfo  {journal} {J. Phys.: Cond. Matt.}\ }\textbf {\bibinfo {volume}
  {20}},\ \bibinfo {pages} {494245} (\bibinfo {year} {2008})}\BibitemShut
  {NoStop}%
\bibitem [{\citenamefont {Fushiki}(1992)}]{Fushiki:1992}%
  \BibitemOpen
  \bibfield  {author} {\bibinfo {author} {\bibfnamefont {M.}~\bibnamefont
  {Fushiki}},\ }\href {\doibase http://dx.doi.org/10.1063/1.463676} {\bibfield
  {journal} {\bibinfo  {journal} {J. Chem. Phys.}\ }\textbf {\bibinfo {volume}
  {97}},\ \bibinfo {pages} {6700} (\bibinfo {year} {1992})}\BibitemShut
  {NoStop}%
\bibitem [{\citenamefont {L\"owen}\ \emph {et~al.}(1992)\citenamefont
  {L\"owen}, \citenamefont {Madden},\ and\ \citenamefont
  {Hansen}}]{Lowen:1992}%
  \BibitemOpen
  \bibfield  {author} {\bibinfo {author} {\bibfnamefont {H.}~\bibnamefont
  {L\"owen}}, \bibinfo {author} {\bibfnamefont {P.~A.}\ \bibnamefont {Madden}},
  \ and\ \bibinfo {author} {\bibfnamefont {J.-P.}\ \bibnamefont {Hansen}},\
  }\href {\doibase 10.1103/PhysRevLett.68.1081} {\bibfield  {journal} {\bibinfo
   {journal} {Phys. Rev. Lett.}\ }\textbf {\bibinfo {volume} {68}},\ \bibinfo
  {pages} {1081} (\bibinfo {year} {1992})}\BibitemShut {NoStop}%
\bibitem [{\citenamefont {L\"owen}\ and\ \citenamefont
  {Kramposthuber}(1993)}]{Lowen:1993}%
  \BibitemOpen
  \bibfield  {author} {\bibinfo {author} {\bibfnamefont {H.}~\bibnamefont
  {L\"owen}}\ and\ \bibinfo {author} {\bibfnamefont {G.}~\bibnamefont
  {Kramposthuber}},\ }\href {http://stacks.iop.org/0295-5075/23/i=9/a=009}
  {\bibfield  {journal} {\bibinfo  {journal} {Eur. Phys. Lett.}\ }\textbf
  {\bibinfo {volume} {23}},\ \bibinfo {pages} {673} (\bibinfo {year}
  {1993})}\BibitemShut {NoStop}%
\end{thebibliography}%
\newpage
\mbox{}
\newpage
\onecolumngrid
\section*{Supplementary Information}
\makeatletter
\renewcommand{\fnum@figure}{\figurename~S1}
\makeatother
\begin{figure*}[h]
\includegraphics[width=\textwidth]{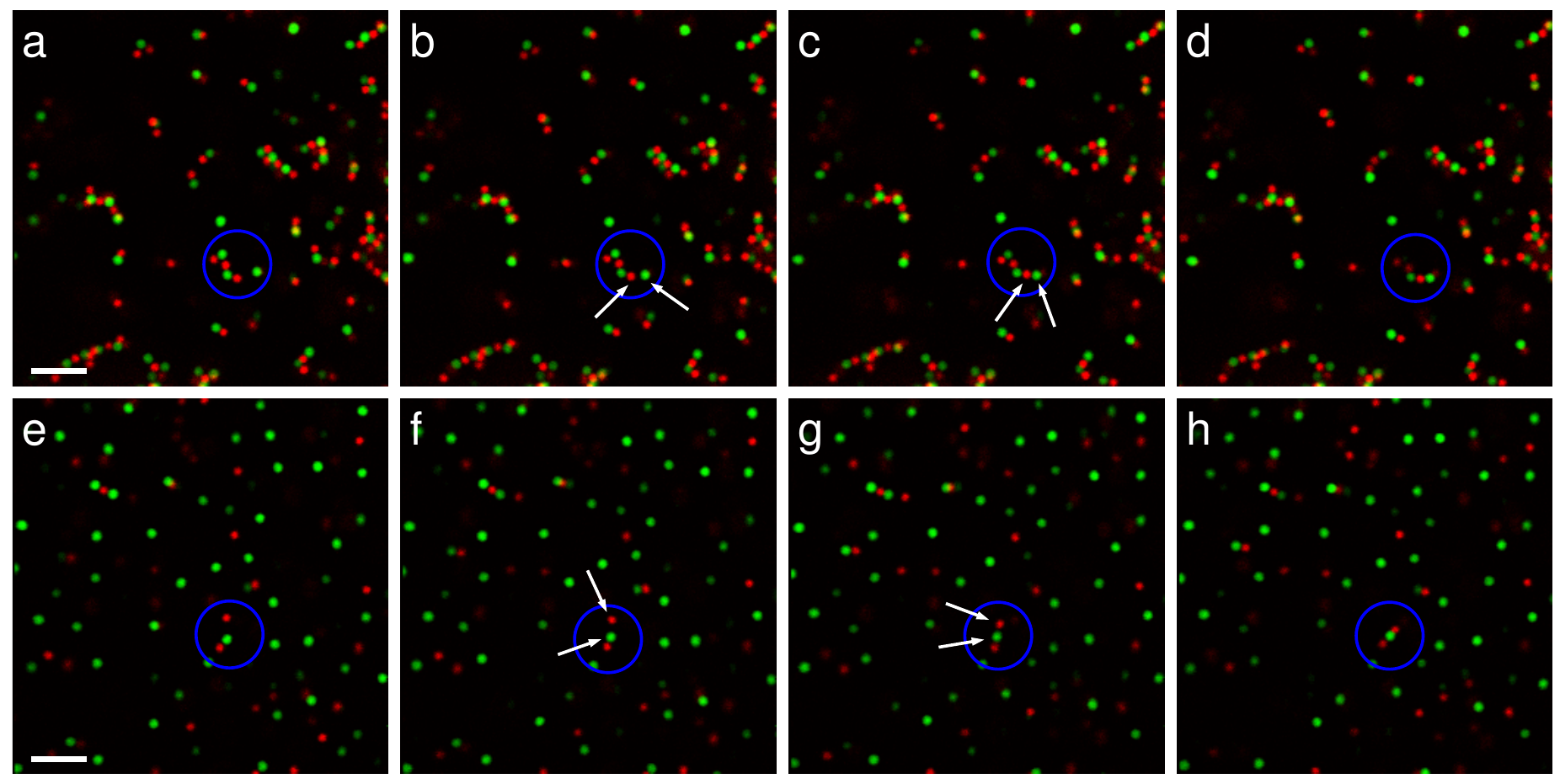}
\caption{An uncropped version of Fig. \ref{CHfig:att}, with scale bars that indicate $10 \ \mu\mathrm{m}$. As in Fig. \ref{CHfig:att}, we show two sequences of four confocal images (a)--(d) and (e)--(h) showing the attachment of a large (green) and small (red) particle. The arrows indicate the particles that are involved in the attachment. The images were taken 1.3 mm and 1.1 mm from the CHB-water interface, $\sim 1$ hours and $\sim 4$ hours after mixing, respectively. In each case we show, from left to right, two images that were taken before attachment, followed by two images taken after attachment. The time step between two consecutive images was 1.04 s.}
\label{fig:noncropped}
\end{figure*}
We also added the following movies that support the main text.
Supplementary movie 1 corresponds to Fig. \ref{CHfig:str}(a) and is shown in real time 5x speeded up. Movie 2 corresponds to Fig. \ref{CHfig:str}(b), also as function of real time, but 7x speeded up. Supplementary movie 3 shows the suspension close to an oil-water interface for various values of the height $z$ (the stacks are made over a total height of 50 $\mu$m), and corresponds to to Fig. \ref{CHfig:str}(c).
Supplementary movie 4 shows the particle attachment (7x speeded up) of Fig. \ref{CHfig:att}(a)-(d) (uncropped Fig. S1(a)-(d)), and a similar attachment process is shown in movie 5 for Fig. \ref{CHfig:att}(e)-(h) (uncropped Fig. S1(e)-(h)). The attachment in both movies occurs after roughly 5 seconds.

\end{document}